\documentclass[a4paper,twocolumn,11pt,accepted=2025-09-23]{quantumarticle}
\pdfoutput=1
\usepackage[utf8]{inputenc}
\usepackage[english]{babel}
\usepackage[T1]{fontenc}
\usepackage{amsmath}
\usepackage{hyperref}
\usepackage{braket}
\usepackage{tikz}
\usepackage{lipsum}
\usepackage{dsfont}
\DeclareMathOperator{\Tr}{Tr}

\newtheorem{result}{Result}

\begin{document}

\title{Certifying quantum enhancements in thermal machines beyond the Thermodynamic Uncertainty Relation}

\author{José A. Almanza-Marrero}
\orcid{0009-0008-1525-5096}
\affiliation{Institute for Cross-Disciplinary Physics and Complex Systems IFISC (UIB-CSIC), Campus Universitat Illes Balears, E-07122 Palma de Mallorca, Spain}
\email{joseantonio@ifisc.uib-csic.es}
\author{Gonzalo Manzano}
\orcid{0000-0003-1359-6344}
\affiliation{Institute for Cross-Disciplinary Physics and Complex Systems IFISC (UIB-CSIC), Campus Universitat Illes Balears, E-07122 Palma de Mallorca, Spain}
\email{gonzalo.manzano@ifisc.uib-csic.es}

\maketitle

\begin{abstract}
  Quantum coherence has been shown to impact the operational capabilities of quantum systems performing thermodynamic tasks in a significant way, and yet the possibility and conditions for genuine coherence-enhanced thermodynamic operation remain unclear. Introducing a comparison with classical machines using the same set of thermodynamic resources, we show that for steady-state quantum thermal machines --- both autonomous and externally driven --- that interact weakly with thermal reservoirs and work sources, the presence of coherence induced by perturbations in the machine Hamiltonian guarantees a genuine thermodynamic advantage under mild conditions. This advantage applies to both cases where the induced coherence is between levels with different energies or between degenerate levels. On the other hand, we show that engines subjected to noise-induced coherence can be outperformed by classical stochastic engines using exactly the same set of (incoherent) resources. We illustrate our results with three prototypical models of heat engines and refrigerators: the three-level amplifier, the three-qubit autonomous refrigerator, and a noise-induced-coherence machine.
\end{abstract}

\section{Introduction} 
The determination of the interplay between quantum coherence ---i.e. the ability of quantum systems to exist in superpositions of multiple states--- and thermodynamic operation, constitutes one of the main challenges in quantum thermodynamics, attracting increasing attention during the last decade~\cite{Binder18,Landi21}. One of the main approaches to exploring this link, in line with the original spirit of thermodynamics, has consisted of the construction and analysis of minimal models of quantum thermal machines showing different types of coherent evolution~\cite{Scovil, Geva94,Boukobza07,Linden10,Levy12,Brunner12,Uzdin15}. These machines consist of a quantum system ---the working medium--- composed of few energy levels or qubits which, by coupling to thermal baths at different temperatures and external work sources, are able to perform useful thermodynamic tasks, such as work extraction, heat pumping, or refrigeration. Recent advances in the manipulation and control of quantum systems in the laboratory allowed the implementation of first prototypes, where the basic principles of these models can be mapped to realistic devices on platforms ranging from ion traps~\cite{Rossnagel16,Maslennikov19} to NV centers in diamond~\cite{Klatzow19}, just to mention a few of them~\cite{Binder18,Deffner22}.

Since the pioneering work of Scully \emph{et al.} introducing a photo-Carnot engine~\cite{Scully03}, quantum coherence has been claimed to increase the power output or efficiency of many different types of quantum heat engines and refrigerators. Such improvements are particularly relevant in the case of continuous machines working in steady state conditions~\cite{Kosloff14}, which in principle require less control of the dynamics and couplings with reservoirs. Particularly relevant examples include power enhancements by noise-induced coherence in lasers, photocell engines, or quantum dots engines~\cite{Scully11,Gelbwaser15,Um22} (with applications in photosynthetic light harvesting~\cite{Dorfman13}), or by input external coherent fields~\cite{Uzdin15,Klatzow19}, as well as cooling boosts by degenerate coherence in local models of quantum absorption refrigerators~\cite{Brunner14,Mitchison15,Manzano19}. In all such cases, coherence has been found to play a positive role in the output mechanism, eventually leading to an increased ability of the machine for work extraction or refrigeration. However, since all such output mechanisms are model dependent, it remains unclear whether the performance shown by these machines cannot be achieved by other equivalent classical models~\cite{Uzdin15,Onam19,Manzano19}, so that a truly quantum thermodynamic advantage can be identified. Given the possibility of implementation of these models in the laboratory and their potential applications, clarifying this point becomes an urgent and crucial point for the field.

\begin{figure*}[t]
  \centering\includegraphics[width=1\textwidth]{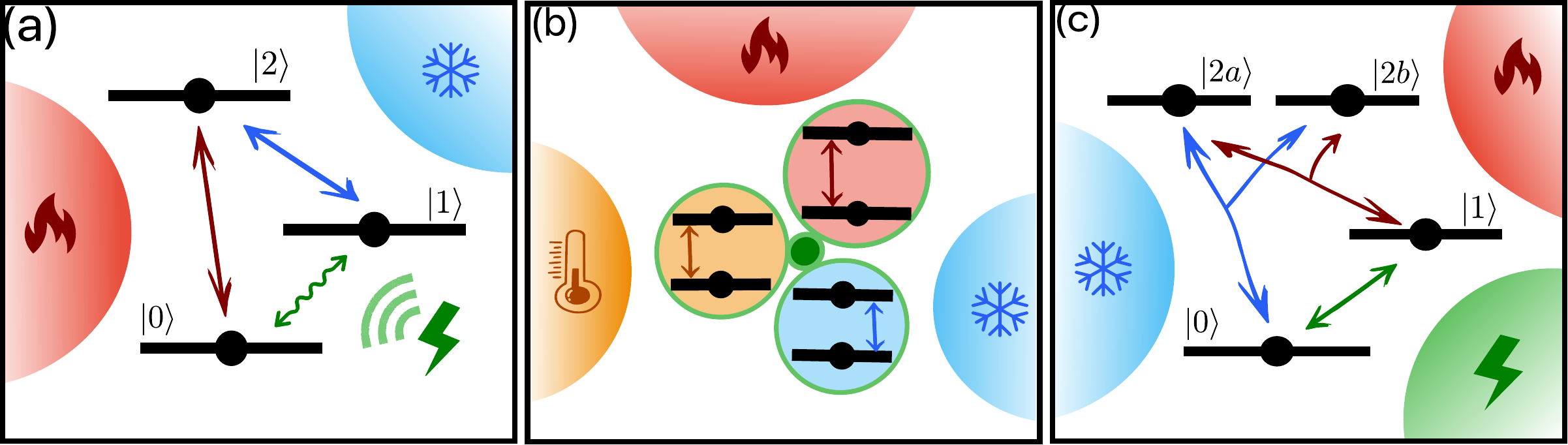}
  \caption{Schematic representation of three quantum thermal machine models. (a) The coherent three-level amplifier with couplings to baths at hot $\beta_\mathrm{h}$(red) and cold $\beta_\mathrm{c}$(blue) temperatures, as well as coherent external driving (green thunderbolt), (b) the three-qubit autonomous (absorption) refrigerator where each qubit is locally coupled to baths at hot (red), cold (blue), and intermediate (yellow) temperatures $\beta_\mathrm{m}$, and (c) the noise-induced-coherence machine showing collective jumps induced by the baths at hot (red) and cold (blue) temperatures, together with a classical work source given by an infinite-temperature bath (green). Plain double arrows represent jumps between the machine energy levels induced by the baths while wavy arrows represent coherent interactions. Collective jumps in (c) are represented by triple arrows.}
  \label{fig:diagrams}
\end{figure*}

In this paper, we show how to identify a genuine coherence-induced quantum thermodynamic advantage and how to quantitatively characterize it in steady-state quantum thermal machines by combining two main ingredients. The first one is the systematic construction of thermodynamically equivalent classical thermal machines that are able to produce the same average currents as their quantum-coherent counterparts, while using exactly the same amount of incoherent resources (essentially bath temperatures and energy structure). However, even if the same average currents can be reproduced by classical models, fluctuations, as captured by the variance of the currents, might present significant differences. The comparison of the fluctuations in the output currents (determining the reliability of the machine) with respect to the classical counterpart is henceforth the second necessary ingredient in our analysis. This is connected with violations of the so-called thermodynamic uncertainty relation (TUR)~\cite{Barato15,Todd16,Horowitz20}, which provides a universal trade-off relation between power, efficiency, and output reliability for any classical Markovian thermal machine operating at steady-state conditions~\cite{Pietzonka18}. The TUR sets up a model-independent limit in the maximum reliability achievable by any classical machine at a given power output and efficiency. Hence, the observation of TUR violations in quantum Markovian machines working in steady-state conditions~\cite{Ptaszy18,Liu19,Kalaee21,Mitchison21,Souza22,Manzano23} may be considered as an unambiguous witness of a quantum thermodynamic signature. However, the contrary is not necessarily true, i.e. the absence of a TUR violation does not imply the absence of thermodynamic enhancements in the machine, as we explicitly show for different classes of quantum machines, like e.g. autonomous multipartite machines. 

Following this method, we certify systematic quantum-thermodynamic advantages in generic steady-state quantum thermal machines (both autonomous or externally driven) that present Hamiltonian-induced coherence in the weak-coupling limit and weak-driving regime. On the other hand, noise-induced coherence can lead to disadvantages in performance, even in cases where the quantum machine dynamics contains intrinsic quantum features.
Moreover, we illustrate our results by employing three well-known (prototypical) models showing coherent-induced evolution (see Fig.~\ref{fig:diagrams}), each corresponding to one of the three possible types of coherence that can arise in the working medium when approaching the steady state: (a) Hamiltonian-induced coherence between different energy levels (energetic coherence) induced by external driving, (b) Hamiltonian-induced coherence between levels with degenerate energies induced by internal couplings, and (c) noise-induced coherence on degenerated levels induced by the reservoir. 

This paper is organized as follows: Sec.~\ref{sec:models} introduces the generic Markovian quantum thermal machine models that we use throughout the manuscript. In Sec.~\ref{sec:equivalents}, we describe the concept of classical equivalent machines and provide a general recipe for constructing them in the presence of different types of coherence. In Sec.~\ref{sec:impact_of_coherence}, we present our findings on the certification of quantum thermodynamic advantages in the form of two main results. 
In Sec.~\ref{sec:illustrative} we verify our results using three specific examples of quantum thermal machines. Finally, Sec.~\ref{sec:conclusions} provides a summary and conclusions of our main results. In the Appendices~\ref{sec:modelsmethods}-\ref{appendix: NICdetails}, we provide details on the particular models employed to test our results, on the characterization of the thermodynamic performance of quantum thermal machines, and detailed proofs on our main results.

\section{Markovian quantum thermal machines models}
\label{sec:models}

 We consider thermal machines running continuously in a steady-state regime, described as open quantum systems that interact with two or more thermal baths at different temperatures. The machine system has $N$ energy levels, some of which may be degenerate or not, and interconnected through incoherent transitions mediated by the baths. Moreover, we consider the possibility of one or more coherent interactions arising as a consequence of one of the three following sources (see Fig.~\ref{fig:diagrams}): (a) external driving by a classical field (such as in quantum heat engine models of masers and lasers~\cite{Scovil,Geva94,Boukobza07}); (b) internal Hamiltonian interactions between subsystems in few-body machines (e.g. machines composed by several interacting qubits~\cite{Linden10} or harmonic oscillators~\cite{Levy12}), and (c) noise-induced coherence caused from collective dissipation acting on two or more resonant transitions (as in some light-harvesting complexes~\cite{Dorfman13} and synthetic heat engine models~\cite{Gelbwaser15}).

The general Hamiltonian of the machine can be generically written as the sum of two terms:
\begin{equation}\label{system_hamiltonian}
    H(t) = H_{0} + V(t), 
\end{equation}
where $H_{0} = \sum_{i=0}^{N-1} \epsilon_{i} \ket{i}\bra{i}$ is a local Hamiltonian describing $N$ energy levels with $\epsilon_0 \leq \epsilon_1 \leq ... \leq \epsilon_{N-1}$, and $V(t)$ is a (eventually time-dependent) term capturing coherent transitions between them. It can either represent an external field periodically driving a transition in $H_0$, or, in the case of multipartite systems, the internal interaction among machine constituents. An important feature of this interaction Hamiltonian is that it does not severely modify the energy level structure of the system, so that it can be treated as a perturbation to the local Hamiltonian $H_{0}$ (i.e. $|V| \ll |H_0|$). On the other hand, for thermal machines using only noise-induced coherence, we will typically have $V(t)=0$ and then $H=H_{0}$, since that type of coherence appears solely from the effect of the baths. Without loss of generality, we set to zero the ground-state energy of the machine, $\epsilon_0 \equiv 0$.

We are interested in the quantum Markovian dynamics of the machine in the weak coupling regime.  
Under Born-Markov and secular approximations, it is possible to describe the evolution of the machine state $\rho(t)$ in terms of a quantum master equation in Lindblad form~\cite{petruccione,book_rivas_huelga, Chiara18, Hofer17,Trushechkin16}:
\begin{equation}\label{quantum_master_equation}
\begin{split}
    \dfrac{d\rho(t)}{dt} = \mathcal{L}(\rho) =& -i \left [ H(t), \rho(t)\right] \\
    &+ \sum_{r=1}^R \sum_{k= \uparrow \downarrow} \mathcal{D}^{(r)}_{k}[\rho(t)],
\end{split}
\end{equation}
where $H(t)$ is given in Eq.~\eqref{system_hamiltonian} and $\mathcal{D}^{(r)}_k(\rho)$ denote the so-called dissipators, taking into account the effects of dissipative process $k$ from thermal reservoir $r$ on the system (we take $\hbar = 1$ through the paper). These dissipators are given in terms of Lindblad operators $L^{(r)}_{k}$ associated to each reservoir:
\begin{equation}\label{dissipator}
\begin{split}
    \mathcal{D}^{(r)}_{k}\left[\rho(t)\right] := L_{k}^{(r)} \rho L_{k}^{(r)~\dagger} - \dfrac{1}{2} \left \{L_{k}^{(r) \dagger}L^{(r)}_{k}, \rho \right \}. 
\end{split}
\end{equation}
The Lindblad operators $L_{k}^{(r)}$ here induce jumps between the $H_0$ levels with fixed energy gap $\Delta \epsilon_k = \pm \Delta \epsilon_r$ determined by bath $r$, and verify $[H_0, L_{k}^{(r)}] = - \Delta \epsilon_k L_{k}^{(r)}$. They can be written in general as:
\begin{equation} \label{eq:Lindblad-ops}
  L_{k}^{(r)} =  \sum_{i,j} \alpha_{i j}^{k} \sqrt{\gamma_{i j}} \ket{j}\bra{i},
\end{equation}
with $\alpha_{i j}^{k} = 1$ if $\epsilon_j - \epsilon_i = \Delta \epsilon_k$ and $0$ otherwise. In the case of non-degenerate transitions, the above operators reduce to simple jumps $L_{k}^{(r)}= \sqrt{\gamma_{ij}} \ket{j}\bra{i}$ between energy levels $i \rightarrow j$. However, the operators in Eq.~\eqref{eq:Lindblad-ops} can also describe collective jumps where two or more transitions with same (degenerate) gap $\Delta \epsilon_k$ in $H_0$ may occur simultaneously, e.g. $i \rightarrow j$ and $i \rightarrow j^\prime$ if $\epsilon_j = \epsilon_j^\prime$. 
In any case,  every transition is connected to a single thermal bath, the rates $\gamma_{ij}\geq 0$ being time independent and verifying the local detailed balance relation $\gamma_{ij} = \gamma_{ji}~e^{-\beta_r (\epsilon_j - \epsilon_i)}$, with $\beta_{r} = 1/{k_B T_r}$ the inverse temperature of the bath $r$ and $k_B$ Boltzmann's constant (in the following we also take $k_B = 1$). In the long-time limit, the evolution dictated by Eq.~\eqref{dissipator} converges to a steady state, verifying $\mathcal{L}(\pi) = 0$, where we denote $\pi(t)$ the steady-state density operator. We notice that, due to the presence of the time-dependent Hamiltonian $V(t)$, the steady state can show a (residual) periodic time dependence in the phase in the Schr\"odinger picture. 

Although the results we present apply to any model of quantum thermal machine whose evolution can be described within the general framework introduced above, we will often particularize to three representative and well-known models of quantum thermal machines, as illustrated in Fig.~\ref{fig:diagrams}. They lead to three different types of coherent evolution respectively: the coherent three-level maser (Fig.~\ref{fig:diagrams}a) which induces energetic coherence in the $H_0$ basis; the three-qubit autonomous quantum refrigerator (Fig.~\ref{fig:diagrams}b), leading to Hamiltonian-induced coherence in a degenerate subspace of $H_0$, and the noise-induced-coherence engine (Fig.~\ref{fig:diagrams}c), showing coherence between degenerate levels induced by collective bath transitions. Details about the three particular models are provided in the Appendix~\ref{sec:modelsmethods} and the thermodynamic performance of generic Markovian quantum thermal machines operating in steady-state conditions is analyzed in Appendix~\ref{sec:performance}.

\section{Classical thermodynamic equivalents of thermal machines}
\label{sec:equivalents}

Throughout this paper, inspired by the notion of classical emulability introduced in Ref.~\cite{Onam19}, we use the term ``classical thermodynamic equivalent'' (or simply ``classical equivalent'') of a quantum thermal machine to refer to a thermal machine model with same bare Hamiltonian, $H_0$, but whose evolution can be described using only classical Markovian dynamics (stochastic jumps between the energy levels), while being capable of producing the same average currents as the quantum machine, by using the same amount of (incoherent) resources. 

By using same incoherent resources, we mean that the classical equivalent model is in contact with the same thermal baths, and therefore has access to the same temperatures. Furthermore, the requirement of having the same bare Hamiltonian $H_0$ implies that the classical equivalent also has the same energy-level structure as the original quantum machine, which allows us to identify them as the ``same machine". This is in contrast with other notions of classical analogs in thermal machines, where all parameters of the classical model are imposed to be exactly equal, including the couplings to the baths or the driving strength~\cite{Uzdin15,Klatzow19} (this is also the case of introducing extra dephasing in the quantum model). Here instead we allow to vary these parameters as long as the requirements for weak-coupling and Markovian dynamics assumed for the thermal machine models are satisfied. This choice is not arbitrary but leads to a stronger notion of genuine quantum-thermodynamic advantage that avoids spurious ``advantages" that may disappear by just slightly modifying some of the parameters in the classical model.

We develop a general method for constructing these equivalent machines in situations where the coherence in the system arises either from Hamiltonian dynamics (Hamiltonian-induced coherence) or from dissipative processes (noise-induced coherence).

\subsection{Hamiltonian-induced coherence}
\label{sec:equivalents.hamiltonian} 
Since we want the classical equivalent to produce the same average energy currents as the quantum counterpart, our starting point will be the generic expressions for the currents, which are detailed in Appendix~\ref{sec:performance}. In particular, the expression for the heat current to reservoir $r$, valid for any given (bare) Hamiltonian $H_0$, can be rewritten as:  
\begin{equation}\label{heat_normal}
    \langle \dot{Q}_{r} \rangle = \sum_{i < j}^{\in B_{r}} \left(\epsilon_{j} - \epsilon_{i}\right) (\gamma_{ij} \pi_{ii} - \gamma_{ji} \pi_{jj}), 
\end{equation}
where we remind the reader that transitions in the sum above are restricted to the ones induced in the set $B_r$, i.e. induced by reservoir $r$ (see Appendix \ref{appendix: general expression heat} for details). The above expression only depends on the level populations (diagonal elements of the density matrix), energy gaps of the machine, and jump rates. As a consequence, the useful output current of the machine can only depend on these quantities as follows from the first law, e.g. in heat engines $\langle \dot{W} \rangle = \sum_r \langle \dot{Q}_r \rangle$. Given that the energy gaps and temperatures of the reservoirs are fixed, we conclude that to ensure identical input and output currents, the quantum machine and its classical equivalent must have matching diagonal elements in their steady-state density matrices.  

In order to mimic the same level populations in the classical equivalent, we proceed by first solving the equations of motion for the diagonal elements and non-vanishing coherences in the quantum machine (details of this procedure are given in App. \ref{appendix: classical equivalent NIC}). Let us denote by indices $\ket{\mathrm{u}}$ and $\ket{\mathrm{v}}$ a couple of levels that are connected by $V$ (for simplicity, we assume $\ket{\mathrm{u}}$ and $\ket{\mathrm{v}}$ not further connected to other levels by $V$):
\begin{equation}\label{eq:driving_exp}
V(t) = g(\ket{\rm u} \bra{\rm v} e^{-i\omega_{\rm d} t} + {\rm h.c.}),    
\end{equation}
with $g \ll \epsilon_{\rm u} - \epsilon_{\rm v}$ and $\omega_{\rm d} = \epsilon_{\rm u} - \epsilon_{\rm v} + \Delta_{\rm d}$ not necessarily resonant with the transition ${\rm u} \leftrightarrow {\rm v}$. If $\epsilon_{\rm u} = \epsilon_{\rm v}$, we recover the case of coherence between degenerate levels. 

The equations of motion for these two levels read in the rotating frame: 
\begin{align}\label{eq._of_Motion}
    \dot{\rho}_{\mathrm{uu}} &= \sum_{j} \gamma_{j \mathrm{u}}\rho_{jj} - \rho_{\mathrm{uu}}\sum_{i}\gamma_{\mathrm{u}i} - 2 g\,\text{Im}\left(\rho_{\mathrm{uv}}\right), \nonumber \\
    \dot{\rho}_{\mathrm{vv}} &= \sum_{j} \gamma_{j\mathrm{v}}\rho_{jj} - \rho_{\mathrm{vv}}\sum_{i}\gamma_{\mathrm{v}i} + 2 g\,\text{Im}\left(\rho_{\mathrm{uv}} \right),\\
    \dot{\rho}_{\mathrm{uv}} &= -\dfrac{1}{2}\sum_{i}\left( \gamma_{\mathrm{u}i} + \gamma_{\mathrm{v} i} - 2 \rm{i} \Delta_d \right) \rho_{\mathrm{uv}} \nonumber \\ 
    &~~~~- \rm{i} g \left ( \rho_{\mathrm{vv}} - \rho_{\mathrm{uu}} \right), \nonumber
\end{align}
while for the rest of levels $n \neq \{\mathrm{u}, \mathrm{v}\}$, not connected by $V$, we simply have $\dot{\rho}_{nn} = \sum_{j \neq n} \gamma_{jn}\rho_{jj} - \rho_{nn}\sum_{i} \gamma_{ni}$.

Following Ref.~\cite{Onam19}, by equating all derivatives to zero in Eqs.~\eqref{eq._of_Motion}, we can determine the relation between the coherence of levels connected by the Hamiltonian $V$ and their populations (which should be verified in the steady state):
\begin{equation}\label{steady_state_coherence}
    \pi_{\mathrm{uv}} = \dfrac{-2  g\left( \pi_{\mathrm{vv}} - \pi_{\mathrm{uu}} \right)}{2\Delta_{\rm d}+\rm i\sum_{j}\left ( \gamma_{\mathrm{u} j} + \gamma_{\mathrm{v} j} \right)}. 
\end{equation}
Then we introduce the above dependence back into off-diagonal elements in Eqs.~\eqref{eq._of_Motion}, to obtain that the net effect of coherence in the steady state is equivalent to adding a virtual transition promoting jumps between the interacting levels $\ket{\mathrm{u}}$ and $\ket{\mathrm{v}}$:
\begin{align}\label{eq._of_Motion_cl}
\begin{split}
    \dfrac{d}{dt} \rho_{\mathrm{uu}} &= \sum_{j\neq \mathrm{u}} \gamma_{j \mathrm{u}}\rho_{jj} - \rho_{\mathrm{uu}}\sum_{i}\gamma_{\mathrm{u}i}  \\ &~~~~~+ \gamma_{\mathrm{vu}}^\text{cl} \rho_{\mathrm{vv}} - \gamma_{\mathrm{uv}}^\text{cl} \rho_{\mathrm{uu}},\\
    \dfrac{d}{dt} \rho_{\mathrm{vv}} &= \sum_{j\neq \mathrm{v}} \gamma_{j\mathrm{v}}\rho_{jj} - \rho_{\mathrm{vv}}\sum_{i}\gamma_{\mathrm{v}i} \\ &~~~~~ + \gamma_{\mathrm{uv}}^\text{cl} \rho_{\mathrm{uu}} - \gamma_{\mathrm{vu}}^\text{cl} \rho_{\mathrm{vv}},
\end{split}
\end{align}
with transition rates
\begin{equation}\label{classical_rate}
    \gamma_{\mathrm{uv}}^{\rm cl} = \gamma_{\mathrm{vu}}^{\rm cl} = \dfrac{4 g^{2}\sum_{i}\left ( \gamma_{\mathrm{u} i} + \gamma_{\mathrm{v} i} \right)}{4 \Delta_{\rm d}^{2}+\left[\sum_{i}\left ( \gamma_{\mathrm{u} i} + \gamma_{\mathrm{v} i} \right)\right]^{2}}.
\end{equation}
As a consequence, if we replace the Hamiltonian term $V$ responsible for the coherent interaction between levels $\ket{\mathrm{u}}$ and $\ket{\mathrm{v}}$ by the above extra stochastic transition between them, the system governed by Eqs.~\eqref{eq._of_Motion_cl} will reach a steady state with exactly the same populations as the original one governed by Eq.~\eqref{eq._of_Motion}, and therefore the same currents in Eq.~\eqref{heat_normal}. 

We have thus achieved a classical equivalent of the quantum model in which the dynamics is entirely classical, as given by a set of incoherent jumps between the machine energy levels, while reproducing the same (average) currents in the steady state. We also notice that the equivalent machine is coupled to the same set of thermal baths, while the coherent interaction is replaced by a stochastic jump process without bias in either direction. This provides a classical equivalent model that uses the same amount of incoherent resources. 
In the case of energetic coherence, i.e. non-degenerate levels $\ket{\mathrm{u}}$ and $\ket{\mathrm{v}}$ ($\epsilon_{\mathrm{u}} \neq \epsilon_{\mathrm{v}}$) this means replacing the quantum-coherent work source (battery) by a classical stochastic work source (battery), like a bath at infinite temperature. On the other hand, in the case of degenerate coherence ($\epsilon_{\mathrm{u}} = \epsilon_{\mathrm{v}}$) the extra stochastic transition is a source of pure noise without any associated energy current and can hence be considered as free.

\subsection{Noise-induced coherence}
\label{sec:equivalents.noise-induced}

In this case, we find that the heat currents of the collective transitions explicitly depend on the real part of the coherence between energy levels involved in the environmental noise-inducing mechanism (which we again denote $\ket{\mathrm{u}}$ and $\ket{\mathrm{v}}$):
\begin{align} \label{eq:NICheat}
    \begin{split}
\langle\dot{Q}_{r}\rangle = &\sum_{i<j}^{\in B_r} (\epsilon_j-\epsilon_i)(\gamma_{ij}\pi_{ii}-\gamma_{ji}\pi_{jj}) \\&+ 2 \sum_{j}^{\in B_r} (\epsilon_j-\epsilon_{\mathrm{v}})\sqrt{\gamma_{\mathrm{u}j}\gamma_{\mathrm{v}j}}~\mathrm{Re}(\pi_{\mathrm{uv}}),
\end{split}
\end{align}
while the currents that are not involved in the noise-inducing mechanism are given by Eq.~\eqref{heat_normal} as in the previous case. Moreover, in Appendix \ref{appendix: classical equivalent NIC}, we show that reproducing the steady-state populations again results in equal average currents, despite the appearance of the second contribution in Eq.~\eqref{eq:NICheat}.

Following the same procedure as for Hamiltonian-induced coherence, we can observe the net effect of coherence in the steady state by solving the equations of motion and replacing the dependence of coherence back into the equations: 
\begin{equation}\label{eq:steady_noise-induced_coherence}
\pi_{\text{uv}} = \dfrac{\sum_{i} \left[ 2 \sqrt{\gamma_{\text{u}i}\gamma_{\text{v}i}}\pi_{ii} - \sqrt{\gamma_{i\text{u}}\gamma_{i\text{v}}} (\pi_{\text{uu}} + \pi_{\text{vv}})\right ]}{\sum_{i}(\gamma_{i\text{u}} + \gamma_{i\text{v}}) }.
\end{equation}
As before, we obtain a virtual jump between the coherent levels $\mathrm{u}$ and $\mathrm{v}$, but in this case we also find corrections to the rates in some of the (already present) jumps involving these levels and other levels of the machine $n$. Thus the rates for the classical equivalent must be of the form: 
\begin{equation}\label{correccion_NI_CA}
\begin{split}
&\gamma_{\mathrm{uv}}^{\mathrm{cl}} = \gamma_{\mathrm{vu}}^{\mathrm{cl}} = \dfrac{\left(\sum_{j}\sqrt{\gamma_{\mathrm{u} j}\gamma_{\mathrm{v} j}}\right)^{2}}{\sum_{j}(\gamma_{\mathrm{u} j}+\gamma_{\mathrm{v} j})},\\
&\gamma_{in}^{\mathrm{cl}} = \gamma_{in} - 2 \sqrt{\gamma_{\mathrm{u} n}\gamma_{\mathrm{v} n}}~\gamma_{\mathrm{uv}}^{*},\\
&\gamma_{ni}^{\mathrm{cl}} = \gamma_{ni} - 2 \sqrt{\gamma_{n \mathrm{u}}\gamma_{n \mathrm{v}}}~\gamma_{\mathrm{uv}}^{*},\\
\end{split}
\end{equation}
where $i= \mathrm{u}, \mathrm{v}$ above and we defined $\gamma_{\mathrm{uv}}^{*} := \sum_{j} \sqrt{\gamma_{\mathrm{u} j}\gamma_{\mathrm{v} j}}/\sum_{j}(\gamma_{\mathrm{u} j}+\gamma_{\mathrm{v} j})$. It can be proved that local detailed balance relations are not modified in any transition of the machine by the corrections to the rates above. To show this, let us rewrite the rates of collective transitions to and from levels $i = \{\mathrm{u}, \mathrm{v}$\} as $\gamma_{i n} = k_{r}^{i} \exp[{\beta_r (\epsilon_i - \epsilon_n)/2}]$ and $\gamma_{ni} = k_{r}^{i}~\exp[{-\beta_r (\epsilon_i - \epsilon_n)/2}]$, respectively, were $k_{r}^i := \sqrt{\gamma_{i n} \gamma_{n i}}$ is a purely kinetic contribution to the rates (not depending on the direction of the jumps), and $\beta_r$ is the temperature of the bath to which the transition is coupled. Using this notation the corrected rates read 
\begin{equation} \label{kinetic}
\begin{split}
    &\gamma_{in}^{\mathrm{cl}} =  \left(k_{r}^{i}-2\gamma_{\mathrm{uv}}^{*}\sqrt{k_{r}^{\mathrm{u}}k_{r}^{\mathrm{v}}}\right) e^{{\beta_r (\epsilon_i - \epsilon_n)/2}} ,\\
&\gamma_{ni}^{\mathrm{cl}} =  \left(k_{r}^{i}-2\gamma_{\mathrm{uv}}^{*}\sqrt{k_{r}^{\mathrm{u}}k_{r}^{\mathrm{v}}}\right) e^{{-\beta_r (\epsilon_i - \epsilon_n)/2}}.\\
\end{split}
\end{equation}
Notice that the corrections affect only the purely kinetic contributions to the rates but not their bias. As a consequence, the classical equivalent employs the same thermodynamic resources (temperatures and energy gaps) as the quantum system. In other words, in order to construct the classical equivalent machine we can replace the collective transitions appearing in the original quantum model by local ones (with a tuned rate) to the same thermal baths, and add an extra stochastic transition between the degenerate levels $\ket{\mathrm{u}}$ and $\ket{\mathrm{v}}$. Such a classical equivalent machine can be defined whenever the rates $\gamma_{i n}^{\rm cl}$ and $\gamma_{i n}^{\rm cl}$ in Eq.~\eqref{correccion_NI_CA} [or equivalently in Eq.~\eqref{kinetic}] are non-negative.

\section{Thermodynamic impact of coherence}
\label{sec:impact_of_coherence}

We are now in a position to analyze and directly compare the performance of quantum thermal machines and their classical thermodynamic equivalent counterparts introduced above to unveil the impact of steady-state coherence (see also App. \ref{sec:performance}). 
We recall that, by construction, the classical equivalent machine reproduces the same average currents in all the transitions, and hence it has the same power and efficiency as the original machine. However, while it is in general desirable for any thermal machine to have a large output current and a high efficiency (low rate of entropy production), in microscopic systems another fundamental factor to consider is the fluctuations associated with the currents (specially in the output power) which characterize the reliability of the thermal machine. The key point of our construction is that the fluctuations in the output current (as captured by the variance), can significantly differ in general between quantum and classical models, making the presence of coherence either beneficial or detrimental for the machine output reliability. 

In order to address the impact of coherence in the reliability of the thermal machines, we introduce the relative difference in the fluctuations between classical and quantum machines as  
 \begin{equation} \label{eq:R}
    \mathcal{R} :=\dfrac{\mathrm{Var}[J_\mathrm{out}^{\mathrm{cl}}] -  \mathrm{Var}[J_\mathrm{out}]}{\mathrm{Var}[J_\mathrm{out}]},
\end{equation}
where $J_\mathrm{out}^{\mathrm{cl}}$ denotes the output current in the classical thermodynamic-equivalent model.
The above ratio measures the relative reduction in the dispersion of the output current in the quantum machine, as compared to the classical one. If $\mathcal{R}<0$, then the classical equivalent provides a more accurate output than the original quantum one; $\mathcal{R}=0$ implies that the quantum and classical models are indistinguishable from their dispersion, and $\mathcal{R}>0$ implies that the output in the quantum machine is more accurate, thus providing a quantum-thermodynamic advantage manifested in an enhanced reliability of the machine for the same average output and efficiency.

Our aim here is to obtain analytical expressions for the variances of the output currents for generic quantum machines and its classical equivalent counterpart by using the full counting statistics formalism. Using it, we construct the relative difference $\mathcal{R}$ in Eq.~\eqref{eq:R} and evaluate its sign. This can be performed in principle for both cases of Hamiltonian-induced coherence and noise-induced coherence. %, following, respectively, Secs.~\ref{sec:equivalents.hamiltonian} and \ref{sec:equivalents.noise-induced}. 
We will first focus on the case of Hamiltonian-induced coherence, since it is in that case that more universal results can be obtained, and then discuss also the case of noise-induced coherence.

As we have seen above, %in Sec.~\ref{sec:equivalents} 
a set of rate equations for the evolution of the populations and coherence is obtained for both the original quantum engine and its classical equivalent counterpart. %, which for Hamiltonian-induced coherence are given in Sec.~\ref{sec:equivalents.hamiltonian}. 
However, we notice that even assuming a single transition with coherence, the general $N$-level problem is composed by $N-2$ equations of the form: \begin{equation}\label{eq:classic_master_eq}
   \dot{\rho}_{nn} = \sum_{j \neq n} (\gamma_{jn}\rho_{jj} - \gamma_{ni} \rho_{nn}) 
%    \dot{p}_i = \sum_{j} \gamma_{ij}p_j,
\end{equation}
plus three more equations for the coherent subspace in the quantum machine, Eqs.~\eqref{eq._of_Motion}, or two for equations for the classical equivalent, Eqs.~\eqref{eq._of_Motion_cl}; with the additional simplification that in steady state conditions, we have $\dot{\rho}_{n n} = 0$ $\forall n$. As a consequence, it is clear that in order to obtain analytical results in the generic case, we need to reduce the dimension of the problem. Here, we are able to reduce (and fix) the dimension of the machine by introducing a ``coarse grained" mesostate, similar to the one presented in Ref.~\cite{Esposito12}, involving almost all of the incoherent transitions as follows.

\begin{figure}[t]
  \centering\includegraphics[width=.97\columnwidth]{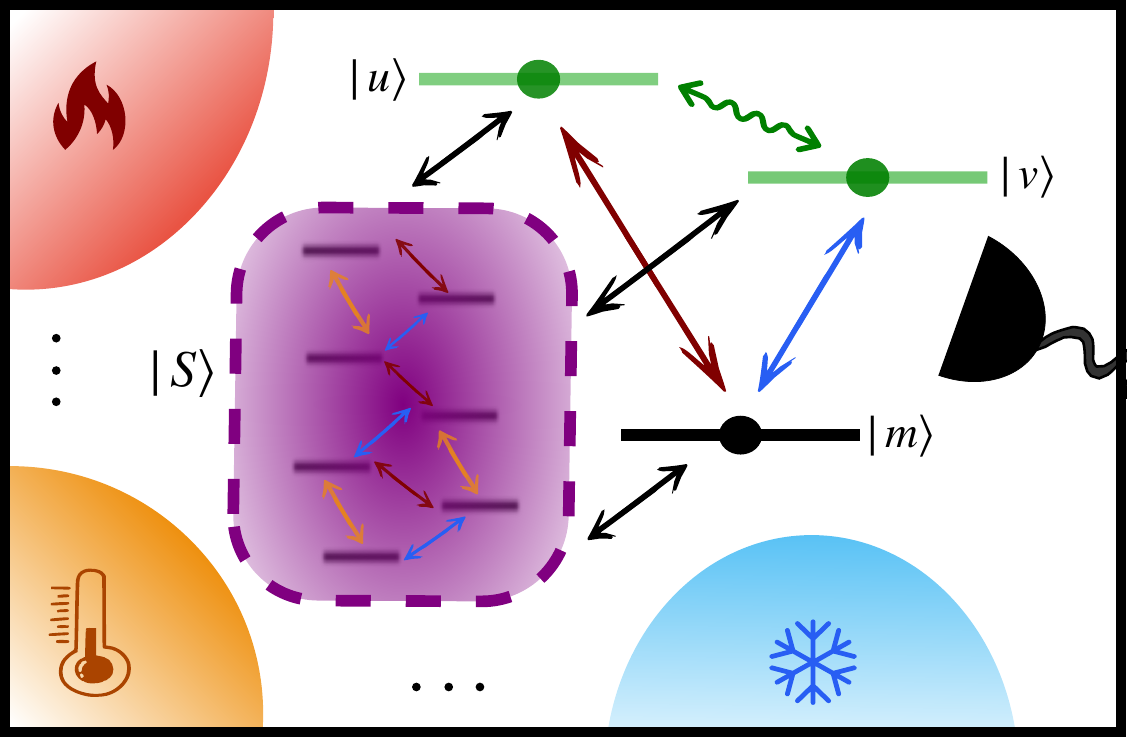}
  \caption{Schematic representation of a generic $N$ level thermal machine with Hamiltonian-induced coherence in a subspace composed by two levels (green levels). The transitions between all other levels are produced by thermal baths at possibly different temperatures (simple arrows), and one of its transitions is being monitored (black detector). A reduced version of the machine model is obtained by introducing the coarse-grained state $S$ (purple shaded box) including $N-3$ levels. In the unicycle case the rates of the transitions $\ket{\rm v}\leftrightarrow \ket{S}$ and $\ket{\rm u} \leftrightarrow \ket{\rm m}$ are set to zero, as well as the transitions between levels within the coarse-grained state $\ket{S}$ leading to multiple cycles.}
  \label{fig: coarse-graining diagram}
\end{figure}

More precisely, we define a ``mesostate" $\ket{S}$ that comprises $N-3$ levels of the thermal machine that are not involved in the coherent interaction (see Fig. \ref{fig: coarse-graining diagram} for a diagram). 
We keep outside $\ket{S}$ only the coherent subspace (levels $\ket{\rm u}$ and $\ket{\rm v}$) and an extra level (denoted without loss of generality as $\ket{\rm m}$ for ``monitoring") connected to it by at least one stochastic transition mediated by a thermal bath (here $\ket{{\rm v}} \leftrightarrow \ket{{\rm m}}$ without loss of generality), that allow us to properly apply the full counting statistics formalism to uncover the fluctuations in the flux over that transition~\footnote{We notice that it is not convenient to apply the FCS formalism to the transitions inside $S$ since, once defined the mesostate, it is not possible anymore to associate an specific energy exchange to any transition from or to $S$.}.

Being composed by an ensemble of levels, the occupation probability of the mesostate can be written as
\begin{equation}
    \rho_{SS} := \sum_{j \in S} \rho_{jj},
\end{equation}
verifying that $\sum_{i\notin S} \rho_{ii} + \rho_{SS} = 1$. Moreover, effective rates for the transitions involving the mesostate and any other level $i$ can be defined as
\begin{equation}
\begin{split}
    &\Gamma_{Si} :=\dfrac{1}{p_S} \sum_{j\in S} \gamma_{ji} \rho_{jj}, ~~;~~ \Gamma_{iS} :=\sum_{j\in S} \gamma_{ij},  
\end{split}
\end{equation}
where only the rates of transitions going ``out" of the mesostate $\Gamma_{Si}$ to the other levels depend on the steady state populations.
Then by introducing the mesostate $S$, the $N-2$ set of equations \eqref{eq:classic_master_eq} can be reduced to the following 2 equations
\begin{align}
    \dot{\rho}_{\rm m m} &= \sum_{i\notin S} \left ( \gamma_{i {\rm m}} \rho_{ii} - \gamma_{{\rm m} i} \rho_{\rm m m} \right ) \nonumber \\ &~~~~+ \Gamma_{S{\rm m}} \rho_{SS} - \Gamma_{{\rm m} S} \rho_{\rm m m}, \\
    \dot{\rho}_{SS} &= \sum_{i\notin S} \left (\Gamma_{i S} \rho_{ii} - \Gamma_{Si} \rho_{SS} \right ). \nonumber \label{eq._coarse_grain}
\end{align}
to be combined with Eqs.~\eqref{eq._of_Motion} for the coherent levels in the quantum machine model, or with Eqs.~\eqref{eq._of_Motion_cl} for the corresponding classical equivalent machine~\footnote{Notice that in this case the sums over levels inside the equations should run over states $\{{\rm u, v, m}, S\}$ with same restrictions as indicated there, and with respective rates.}. In any case, these extra equations remain unchanged by substituting the remaining levels with mesostate $\ket{S}$.

Notably, with the above procedure we have fixed the dimension of the general problem to a 4-states thermal machine with coherence between two of these states. That means applying FCS with a matrix of dimension 5 in the quantum case and of dimension 4 for the classical thermodynamic equivalent machine (see App. \ref{appendix: cumulants}). Here, it is also important to remark that, by allowing transitions between $\ket{S}$ and all the other (three) states of the machine (as well as transitions among the levels within $S$), this general procedure applies to thermal machines that can be unicycle or multicyclic, where the unicycle case is recovered by setting to zero the rates of the transition $\ket{{\rm v}} \leftrightarrow \ket{S}$ and $\ket{\rm u} \leftrightarrow \ket{\rm m}$.

The next step is to obtain analytical expressions for the variance of the current in the monitored transition $\ket{\rm v} \leftrightarrow \ket{\rm m}$. This is done for both the quantum machine and its classical counterpart. Here, it is important to mention that, in order to assess the fluctuations of relevant steady-state currents (including coherent work in the transition $\ket{\rm u} \leftrightarrow \ket{\rm v}$) from FCS in a single machine transition, we have to impose some specific conditions on the transition structure of the machine.
In particular, we assume that every transition in the machine is coupled (at most) to a single thermal reservoir. Moreover, in the case of multicyclic machines, 
we impose that for every single cycle of the original machine involving the coherent transition, the same net number of quanta is exchanged with the reservoirs to complete the cycle in a given direction (or minus that number in the opposite direction). Otherwise, for eventual cycles not involving the coherent transition, the net number of quanta exchanged with all the reservoirs must be zero. This condition can be considered as a version of the tight coupling condition at the level of fluctuations. It (trivially) includes uniclycle machines, but also a broad class of multicycle structures that we dub ``symmetric multiclyce" machines. A relevant example of those structures is obtained in multipartite scenarios such as the three-qubit refrigerator in Fig.~\ref{fig:diagrams}b. There all cycles involve the net exchange of one quantum with each reservoir when they are closed using the coherent transition (energy transfer between the qubits), or zero otherwise (i.e. when qubits just exchange quanta back and forth with their local reservoirs). This is also the case for many other engines and refrigerators, such as e.g. continuous versions of the SWAP engine~\cite{Campisi15}, fridges made up of bosonic modes~\cite{Levy12,Maslennikov19} or using optomechanical-like couplings~\cite{Naseem2020}, as well as quantum dot devices~\cite{Venturelli13,Thierschmann15}, just to mention a few.

In all these situations, the variances of the currents are proportional to each other, ${\rm Var}[\dot{Q}_r] = \Delta\epsilon_r^2 {\rm Var}[\dot{N}]$ for all reservoirs $r$ with ${\rm Var}[\dot{N}] := {\rm Var}[\dot{N}_l]~ \forall l$, and also ${\rm Var}[{\dot{W}}]= (\epsilon_{\rm u} - \epsilon_{\rm v})^2 {\rm Var}[\dot{N}]$ for the power output (see App. \ref{appendix: full counting}). Therefore, the relative difference in the fluctuations in Eq.~\eqref{eq:R} can be conveniently rewritten as:
\begin{equation}
    \mathcal{R} :=\dfrac{\mathrm{Var}[\dot{N}^{\mathrm{cl}}] -  \mathrm{Var}[\dot{N}]}{\mathrm{Var}[\dot{N}]},
\end{equation}
where ${\rm Var}[\dot{N}^{\rm cl}]$ denotes the corresponding variance of the current in the classical equivalent model. The explicit expressions for ${\rm Var}[\dot{N}]$,  ${\rm Var}[\dot{N}^{\rm cl}]$ are provided in Appendix \ref{appendix: cumulants} for a generic quantum machine with $N$ levels and a coherent transition. In the case of several (independent) coherent transitions, they can be handled by using the same techniques independently for every coherent subspace, all of which are kept outside the coarse-grained state $S$ (hence augmenting the effective dimension of the system as a counterpart).

Once the general expressions for the variances are obtained, the sign of $\mathcal{R}$ can be evaluated to observe its universal character. We summarize this analysis in the following result, whose explicit proof is provided in Appendix \ref{appendix: proofs}:  

\begin{result} \label{th1}
    Under the presence of Hamiltonian-induced coherence, quantum thermal machines outperform the precision of their classical thermodynamic equivalent counterparts  in steady-state conditions under weak coupling and weak resonant driving. In particular, it is verified that the relative difference in the fluctuations is always strictly positive $\mathcal{R}>0$ when the systems are in out-of-equilibrium conditions and becomes zero in equilibrium. For non-resonant driving, quantum precision improvements require a bounded detuning $|\Delta_\mathrm{d}| < \sum_{i}\left ( \gamma_{\mathrm{u} i} + \gamma_{\mathrm{v} i} \right)/2$. 
\end{result} 

The above result means that, out of equilibrium, the quantum thermal machine is always more precise than its classical thermodynamic equivalent. Since the classical thermodynamic equivalent will have by construction equal average output currents and efficiency than the quantum machine, that implies that Hamiltonian-induced coherence invariably leads to quantum-thermodynamic advantages. In other words, there is no classical model that, using the same resources, can reproduce the power, efficiency, and reliability of the original quantum thermal machine. 

Result~\ref{th1} constitutes the main outcome of our fluctuation analysis, applicable to generic uni-cyclic and symmetric multi-cycle quantum thermal machines operating in nonequilibrium stationary states (NESS) in the weak coupling regime and under weak driving or perturbations, where the dynamics can be described by the Lindblad master equation, Eq.~\eqref{quantum_master_equation}. Examples of such thermal machines are ubiquitous~\cite{Binder18}, comprising few level masers and lasers under weak driving~\cite{Li17,Klatzow19,Zou17,Zhang22}, superconducting devices~\cite{Hofer16b,Sanchez18,Manzano23}, or many-body engines~\cite{Carollo2020,Souza22}, as well as a number of autonomous engines, refrigerators, and quantum clock models~\cite{Venturelli13,Brask15,Hofer16,Mitchison16,Erker17,Haack21} composed by various units that interact weakly among them.

For deriving Result~\ref{th1}, we assumed a single coherent transition in the thermal machine. However, in situations with several coherent transitions that do not share the same energy levels, a similar reasoning can be applied. In particular, in such cases one can handle one of the coherent transitions at a time, by constructing a series of intermediate classical equivalent machines following the above recipe. In that way, one can first replace the first coherent transition by a classical jump at the corresponding rate dictated by our results. That intermediate machine produces the same average fluxes, but with reduced fluctuations under the conditions of Result~\ref{th1}. Then, one can replace the second coherent transition, obtaining a new classical equivalent with now two of the coherent transitions replaced by classical ones. That procedure can then be repeated further until all the coherent transitions have been replaced. In each of these intermediate equivalent machines the fluctuation ratio of the output currents will be reduced while maintaining the same average currents, thus ensuring a quantum thermodynamic advantage.

\begin{figure*}[t]
  \includegraphics[width=1.0\textwidth]{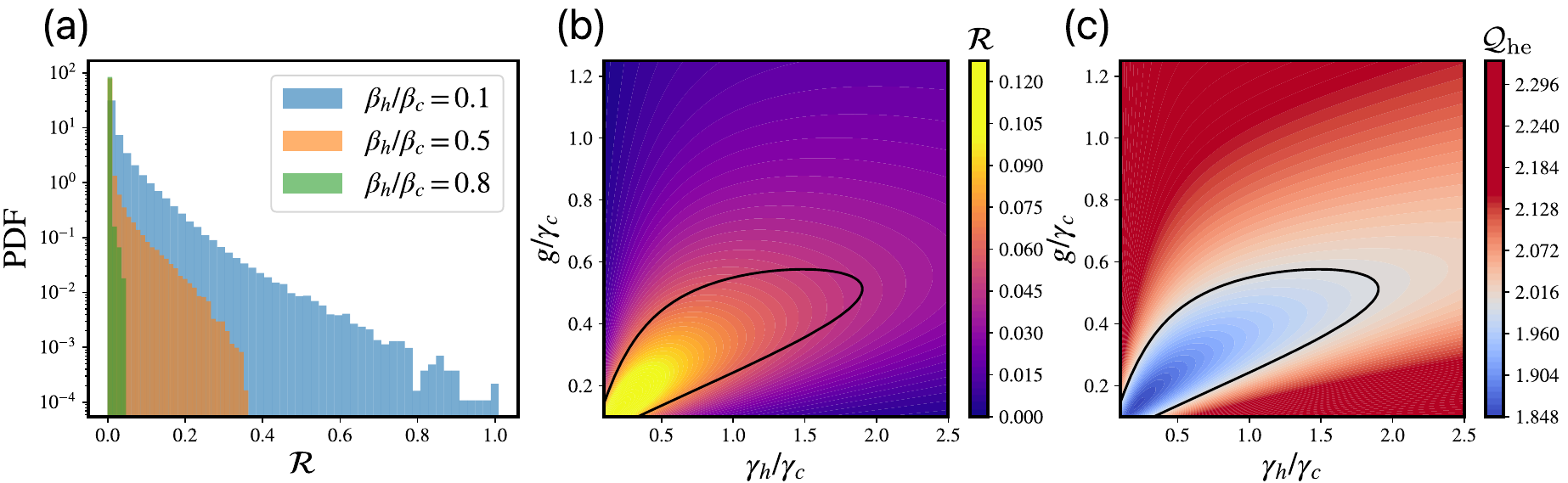}
  \caption{Testing fluctuations in the three-level amplifier. {a}) Histogram of sampled values of the ratio between the fluctuations of the system and its classical analog $\mathcal{R}$ for different ratios of the bath temperatures (see legend). The values corresponds to an exploration of the following region in the parameters space of the system: $\beta_{\mathrm{c}} = 1$, $\epsilon_{2} = 5$, $\omega_{\mathrm{d}} \in [0.1,4.9]$, $\gamma_{\mathrm{h}/\mathrm{c}} \in [10^{-5}, 10^{-2}]$ and $g \in [10^{-5},10^{-2}]$. b) Colour maps of the fluctuation ratio $R$ and c) the TUR ratio $\mathcal{Q}_\mathrm{he}$ as a function of the bath interaction strengths and the driving force. Solid black lines in both plots correspond to the TUR ratio saturation $\mathcal{Q}_{\rm he} = 2$. The other systems parameters are: $\beta_c = 1$, $\beta_{\mathrm{h}} = 0.1/\beta_c$, $\epsilon_2 = 5$, $\omega_{\mathrm{d}} = 2.5$ and $\gamma_{\mathrm{c}} = 10^{-3}$ (energies are given in $k_{\rm B} T_{\rm c} = 1$ units). }
  \label{fig:maser results}
\end{figure*}

The impact of coherence in the case of noise-induced coherence can be analyzed using a similar method as in the Hamiltonian case. However, in this case, more complications arise. The set of rate equations describing the evolution of the system is given in the Appendix \ref{appendix: classical equivalent NIC}.  
In that case, a mesostate similar to the one above might be introduced, but it can only include all the machine levels that are not directly connected to the coherent subspace spanned by states ${\rm u}$ and ${\rm v}$. All other state transitions will be now affected by the presence of the coherent transition [c.f. Eq.~\eqref{correccion_NI_CA}] and the allowed parameter regimes need to ensure their positivity. Moreover, the intrinsic multiclycic structure no longer ensures that the variances of the steady-state currents in all the transitions are equal as before, but the results may depend on the chosen level to perform the FCS monitoring. 

As a consequence, contrary to the Hamiltonian-induced case, the impact of noise-induced coherence turns out to be not universal and depends on the specific model, the specific parameters of the machine, and the specific transition in which FCS is performed, leading to the following: 
\begin{result} \label{th2}
    Under the presence of noise-induced coherence, quantum thermal machines in the weak coupling limit can either outperform or be outperformed by their classical thermodynamic equivalent counterparts in steady-state conditions. In particular, the relative difference in the fluctuations $\mathcal{R}$ can be either positive or negative depending on the specific parameters of the machine.
\end{result} 
The proof of this result is provided in the next section and follows from constructing and analyzing a (counter) example, the NIC machine in Fig.~\ref{fig:diagrams}c, where regimes with $\mathcal{R}>0$ and $\mathcal{R}<0$ are founded for different parameter regimes. The importance of Result~\ref{th2} is based on the fact that it places firmer criteria for the classification of quantum thermodynamic advantages in noise-induced machines than the ones previously considered in the literature, such as comparing with a dephased model~\cite{Scully11,Gelbwaser15,Um22, Dorfman13}. At the same time, it goes beyond the absence of TUR violations, a fact that has been observed in some noise-induced models, see e.g. Ref.~\cite{Segal21}.

\section{Illustrative examples}
\label{sec:illustrative}

We now illustrate our results for the three paradigmatic examples of quantum machines in Fig.~\ref{fig:diagrams} constructing explicitly their classical equivalent machines and assessing the role of coherence in the fluctuations of the currents. We numerically generate a $10^{6}$ number of possible system configurations in the whole parameter space that verify the basic assumptions that ensure the consistency of the Markovian dynamics and perform a direct comparative analysis between quantum and classical models.

\subsection{Three level amplifier}

Following the general recipe, the classical equivalent of the three-level amplifier can be obtained by replacing the driving Hamiltonian $V(t)$ by an extra stochastic transition between levels $\ket{0}$ and $\ket{1}$. The rates of these extra transitions, as given by Eq.~\eqref{classical_rate} become $\gamma_{01}^{\mathrm{cl}} = \gamma_{10}^{\mathrm{cl}} = {4g^2}/{(\gamma_{\mathrm{h}}\bar{n}_{\mathrm{h}} + \gamma_{\mathrm{c}}\bar{n}_{\mathrm{c}})}.$

In order to quantify the impact of (energetic) coherence in the three-level amplifier, we compute the fluctuations ratio $\mathcal{R}$ in Eq.~\eqref{eq:R} for the output power, $J = \dot{W}$. Exploring the model parameters with fixed external temperatures of the baths, we observe the appearance of a significant amount of configurations with $\mathcal{R}>0$, which increases as the temperature bias powering the machine increases. This is illustrated in Fig.~\ref{fig:maser results}{a} where the distribution of $\mathcal{R}$ values over $10^6$ configurations is shown for three different choices of (fixed) environmental temperatures. As can be appreciated, for all configurations we obtain $\mathcal{R}\geq 0$, that is, the three-level amplifier always matches or exceeds the performance of the corresponding classical equivalent machine for each configuration, thus unveiling a beneficial role of energetic coherence. Improvements reaching a reduction in the variance of the output power up to $\mathcal{R} \sim 1$ are possible for temperature bias of the order $T_\mathrm{h} = 10 T_\mathrm{c}$. We also observe a fat tail in the distribution that ensures the robustness of the enhancements, meaning that many configurations can lead to significant reductions in the output variance. The range of variance reductions shrinks towards higher $\mathcal{R}$ values as the temperature bias is reduced, and tends to disappear close to equilibrium (similar temperatures of the baths) where quantum and classical models perform almost equally. This effect is a manifestation of the nonequilibrium character of the enhancements produced by energetic coherence.  

In Fig.~\ref{fig:maser results}{b} the behavior of the fluctuations reduction ratio, $\mathcal{R}$ in Eq.~\eqref{eq:R}, is plotted as a function of the spontaneous emission rates and the driving strength. Darker colors denote regions where larger stability enhancements with respect to the classical equivalent machine are obtained, which are verified in the regime of very weak driving and highly asymmetric spontaneous rates (low coupling strength with the hot bath as compared to the cold one). This plot can be contrasted with Fig.~\ref{fig:maser results}c where the TUR ratio $\mathcal{Q}_\mathrm{he}$ [Eq.~\eqref{eq:Qengine}] is shown for the same range of parameters. In both plots, the black solid line has been introduced as a guide to the eye that represents the boundary ($\mathcal{Q}_\mathrm{he} = 2$) of the region where TUR violations are obtained. This allows a comparison of the method using the classical equivalent machine to detect quantum thermodynamic enhancements, with the direct search for violations of the (classical) TUR~\cite{Ptaszy18,Liu19,Kalaee21,Mitchison21,Souza22,Manzano23}. 

We observe that the area where violations of the TUR, $\mathcal{Q}_\mathrm{he} <2$, are verified, is contained within the region $\mathcal{R}>0$, and it indeed coincides with the highest precision improvements, measured by the reduction ratio $\mathcal{R}$. However, as expected, we also find that even in regimes where the TUR is not violated, there exists an improvement in accuracy of the output current due to the presence of coherence. Therefore, using the classical equivalent of the original three-level amplifier, we can identify regimes of thermodynamic enhancement that cannot be revealed by violations of the TUR. 

\subsection{Three-qubit autonomous refrigerator}

For the autonomous refrigerator model, we find that the classical equivalent is obtained by replacing the three-body interaction Hamiltonian $V$, allowing the exchange of energy between qubits by a classical transition, producing incoherent jumps between levels $\ket{101}$ and $\ket{010}$. The rate of this transition, according to Eq.~\eqref{classical_rate} becomes in this case $\gamma_{\mathrm{uv}}^{\mathrm{cl}} = \gamma_{\mathrm{vu}}^{\mathrm{cl}} = {4g^2}/\sum_{r}{(2\bar{n}_{r}+1)\gamma_{r}}$, where $\ket{\mathrm{u}} = \ket{101}$ and $\ket{\mathrm{v}} = \ket{010}$, and the sum runs over the three baths, $r = \mathrm{h,m,c}$.

\begin{figure}[t]
  \centering\includegraphics[width=1\columnwidth]{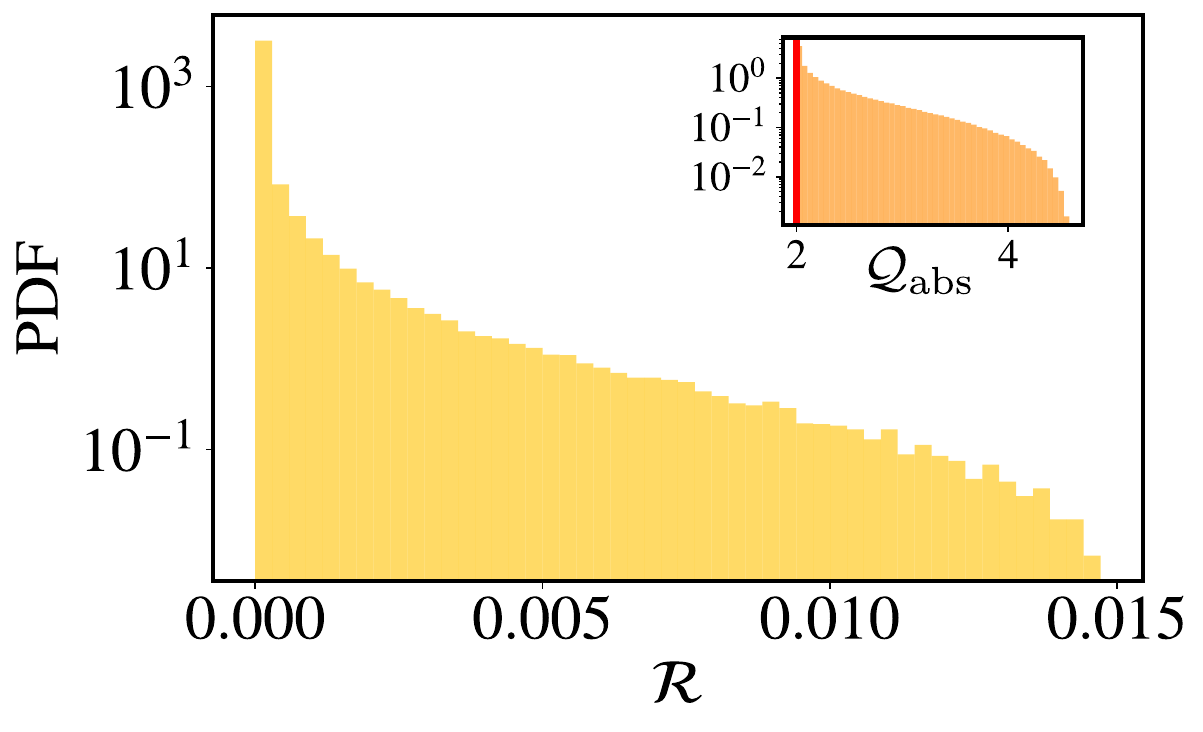}
  \caption{Histograms for the three-qubit refrigerator of sampled values of the fluctuation ratio $\mathcal{R}$ and TUR ratio $\mathcal{Q}_{\rm abs}$ (inset plot). The positive values of $\mathcal{R}$ indicate reduced fluctuations in the quantum model with respect to the classical equivalent that is not related with a violation of the TUR. The values corresponds to an exploration of the following region in the parameters space of the system: $\beta_{\mathrm{m}}/\beta_{\mathrm{c}} \in [0,1]$, $\beta_{\mathrm{h}}/\beta_{\mathrm{m}} \in [0,1]$,  $\epsilon_{1} \in [0.1,4.9]$, $\gamma_{\mathrm{c}/\mathrm{m}/\mathrm{h}} \in [10^{-5}, 10^{-2}]$ and $g \in [10^{-4},10^{-2}]$. Fixed parameter are $\beta_{\mathrm{c}} = 1$, $\epsilon_{2} = 5$ (energies are given in $k_{\rm B} T_{\rm c} = 1$ units).}
  \label{fig:fridges results 1}
\end{figure}

To evaluate the improvements in the reliability of the refrigerator, the relative fluctuations $\mathcal{R}$ in Eq.~\eqref{eq:R} are calculated for the cooling power (heat current from the cold bath), $J_\mathrm{out} = \dot{Q}_\mathrm{c}$. The distributions for the reduction ratio $\mathcal{R}$ and TUR ratio $\mathcal{Q}_\mathrm{abs}$ [Eq.~\eqref{eq:Qabs}] (inset) in this case are shown in Fig.~\ref{fig:fridges results 1} again for $10^6$ parameter configurations. 
As can be observed, in this case, we obtain $\mathcal{R}\geq 0$ for all configurations, meaning that a reduction of the fluctuations ratio is always possible. As a consequence, we have to conclude that the three-qubit quantum refrigerator performs better than its classical equivalent, in alignment with previous works showing that it operates using entanglement~\cite{Brunner14}. 
However, in most cases, the fluctuations in both systems are comparable, with certain parameter regimes where the quantum system exhibits noise levels up to $1.5\%$ lower than its classical thermodynamic equivalent. Looking at the distribution of $\mathcal{Q}_\mathrm{abs}$ values, we also see that the autonomous refrigerator remains unable to break the classical constraint imposed by the TUR in all system configurations (inset plot).

\subsection{Noise-induced coherent machine}
\label{sec:illustrativeC}
In the case of the NIC machine, previous studies have shown that when the rates of the collective transitions were equal ($\gamma_{na} = \gamma_{nb}$ and $\gamma_{an} = \gamma_{bn}$ for $n=0,1$) it is possible to apply a change of variables that effectively decouples coherences from populations~\cite{Holubec18} (see also Ref.~\cite{Manzano18} for a similar case). However, such a change of variables does not produce this decoupling when the rates are unequal, suggesting that the system enters a purely quantum regime~\cite{Holubec19}. In the following, we show that in both cases a classical equivalent can be defined for large regions of the parameter space, which can be used to evaluate the thermodynamic impact of noise-induced coherence.
 
\begin{figure}[t!]
  \includegraphics[width=1\columnwidth]{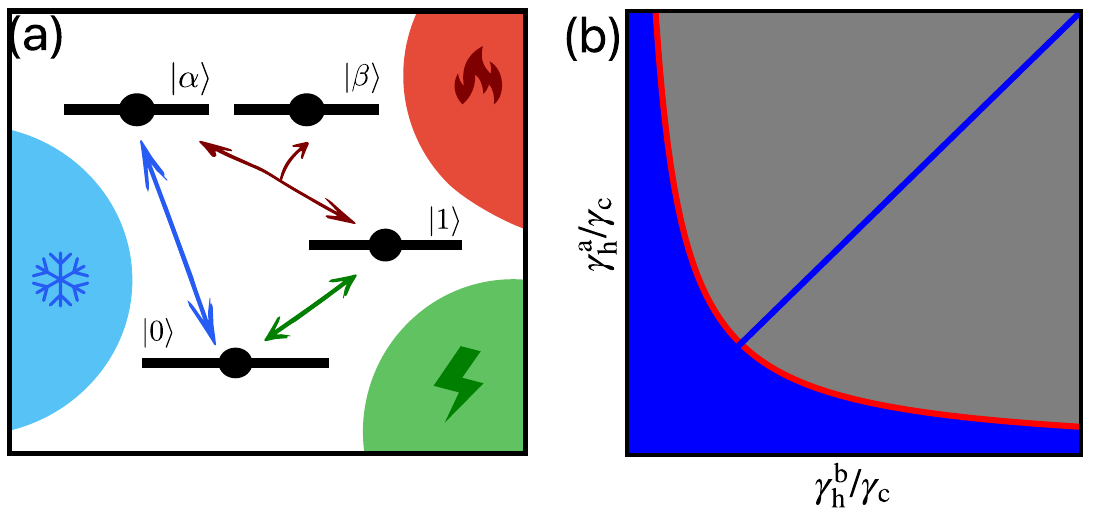}
  \caption{(a) Schematic representation of the level transitions in the NIC machine after the change of variables. Here one of the double coherent transitions are replaced by a simple jump ($\ket{0} \leftrightarrow \ket{\alpha}$). (b) Graphical representation of the inequality (\ref{eq: validity of the equivalent}) for $\gamma^{\rm a}_{\rm c} = \gamma^{\rm b}_{\rm c} =: \gamma_{\rm c}$, separating the regimes where we can define the classical equivalent for noise-induced coherence (blue zone) and where we cannot (grey zone). In the symmetric case (blue diagonal) the classical equivalent can always be defined.}
  \label{fig:regimes_of_equivalence}
\end{figure}

We first introduce the following basis change within the coherent subspace, where the degenerate levels $\ket{2a}$ and $\ket{2b}$ are transformed into levels $\ket{\alpha}$ and $\ket{\beta}$, with: 
\begin{equation}\label{eq: variable_change}
\begin{split}
    &\ket{\alpha} = \dfrac{1}{\sqrt{\gamma_{\mathrm{c}}^{\mathrm{a}} + \gamma_{\mathrm{c}}^{\mathrm{b}} }} \left (\sqrt{\gamma_{\mathrm{c}}^{\mathrm{a}} }\ket{\mathrm{2a}} + \sqrt{\gamma_{\mathrm{c}}^{\mathrm{b}} }\ket{\mathrm{2b}} \right),\\
    &\ket{\beta} = \dfrac{1}{\sqrt{\gamma_{\mathrm{c}}^{\mathrm{a}} + \gamma_{\mathrm{c}}^{\mathrm{b}}} } \left (\sqrt{\gamma_{\mathrm{c}}^{\mathrm{b}} }\ket{\mathrm{2a}} - \sqrt{\gamma_{\mathrm{c}}^{\mathrm{a}}} \ket{\mathrm{2b}} \right). 
\end{split}
\end{equation} 
Notice that in the case of symmetric rates for the cold reservoir, $\gamma_{\mathrm{c}}^{\mathrm{a}} = \gamma_{\mathrm{c}}^{\mathrm{b}}$, this expression becomes the one presented in Ref.~\cite{Holubec19}, but differs otherwise. By introducing this change, we successfully decouple the state $\ket{\beta}$ from the state $\ket{0}$, resulting in a machine with only one explicit collective transition, as illustrated in Fig. \ref{fig:regimes_of_equivalence} (see App. \ref{appendix: NICdetails} for more details). 

The classical equivalent of the NIC machine includes an extra stochastic transition between the degenerate levels reading:
\begin{equation}\label{eq:rates_noise-induced}
\begin{split}
     &\gamma_{\alpha\beta}^{\mathrm{cl}} = \dfrac{(\bar{n}_{\mathrm{h}}+1)^2\gamma_{\mathrm{h}}^{\alpha}\gamma_{\mathrm{h}}^{\beta}}{(\bar{n}_{\mathrm{c}}+1)\gamma_{\mathrm{c}}^{\alpha}+ (\bar{n}_{\mathrm{h}}+1)(\gamma_{\mathrm{h}}^{\alpha}+\gamma_{\mathrm{h}}^{\beta})}, \\
\end{split}
\end{equation}
and $\gamma_{\beta\alpha}^{\mathrm{cl}} = \gamma_{\alpha\beta}^{\mathrm{cl}}$, together with corrections to the rates of the collective transitions  
\begin{equation}
\begin{split} \label{eq:newrates}
    \gamma_{k 1}^{\mathrm{cl}} &= \left(\gamma_{\mathrm{h}}^{k} - \dfrac{2\gamma_{\alpha\beta}^{\mathrm{cl}}}{\bar{n}_{\mathrm{h}}+1}\right) (\bar{n}_{\mathrm{h}}+1), \\
    \gamma_{1 k}^{\mathrm{cl}} &=  \left(\gamma_{\mathrm{h}}^{k} - \dfrac{2\gamma_{\alpha\beta}^{\mathrm{cl}}}{\bar{n}_{\mathrm{h}}+1}\right) \bar{n}_{\mathrm{h}},
\end{split}
\end{equation} 
for $k= \alpha, \beta$.
By examining the above equations, we find that the corrections to the spontaneous emission rates could make these rates eventually negative, which would correspond to a non-physical situation. Consequently, the construction of the classical equivalent for noise-induced coherence is limited to scenarios that ensure positive rates in Eq.~\eqref{eq:newrates}. For the NIC machine analyzed here, this happens when
\begin{equation}\label{eq: validity of the equivalent}
\begin{split}
    \gamma_{\mathrm{h}}^{\beta}=0;&\;\;\;\gamma_{\mathrm{h}}^{\alpha}>0,\\
\gamma_{\mathrm{h}}^{\beta}>0;\;\;\;\gamma_{\mathrm{h}}^{\alpha}>\gamma_{\mathrm{h}}^{\beta};&\;\;\;\dfrac{(\gamma_{\mathrm{h}}^{\alpha}-\gamma_{\mathrm{h}}^{\beta})}{\gamma_{\mathrm{c}}^{\alpha}}\leq \dfrac{\bar{n}_{\mathrm{c}}+1}{\bar{n}_{\mathrm{h}}+1},\\\gamma_{\mathrm{h}}^{\beta}>0;\;\;\;\gamma_{\mathrm{h}}^{\alpha}\leq\gamma_{\mathrm{h}}^{\beta};&\;\;\;\dfrac{(\gamma_{\mathrm{h}}^{\beta}-\gamma_{\mathrm{h}}^{\alpha})}{\gamma_{\mathrm{c}}^{\alpha}}\leq \dfrac{\bar{n}_{\mathrm{c}}+1}{\bar{n}_{\mathrm{h}}+1},\\
\end{split}
\end{equation} 
which, taking into account the variable change together with $\bar{n}_{\mathrm{h}}\geq\bar{n}_{\mathrm{c}}$, corresponds to the stimulated emission rate in the cold bath being greater than or equal to the difference in stimulated emission rates in the hot bath between transitions from levels $\ket{\alpha}$ and $\ket{\beta}$.

\begin{figure}[t!]
  \centering\includegraphics[width=1\columnwidth]{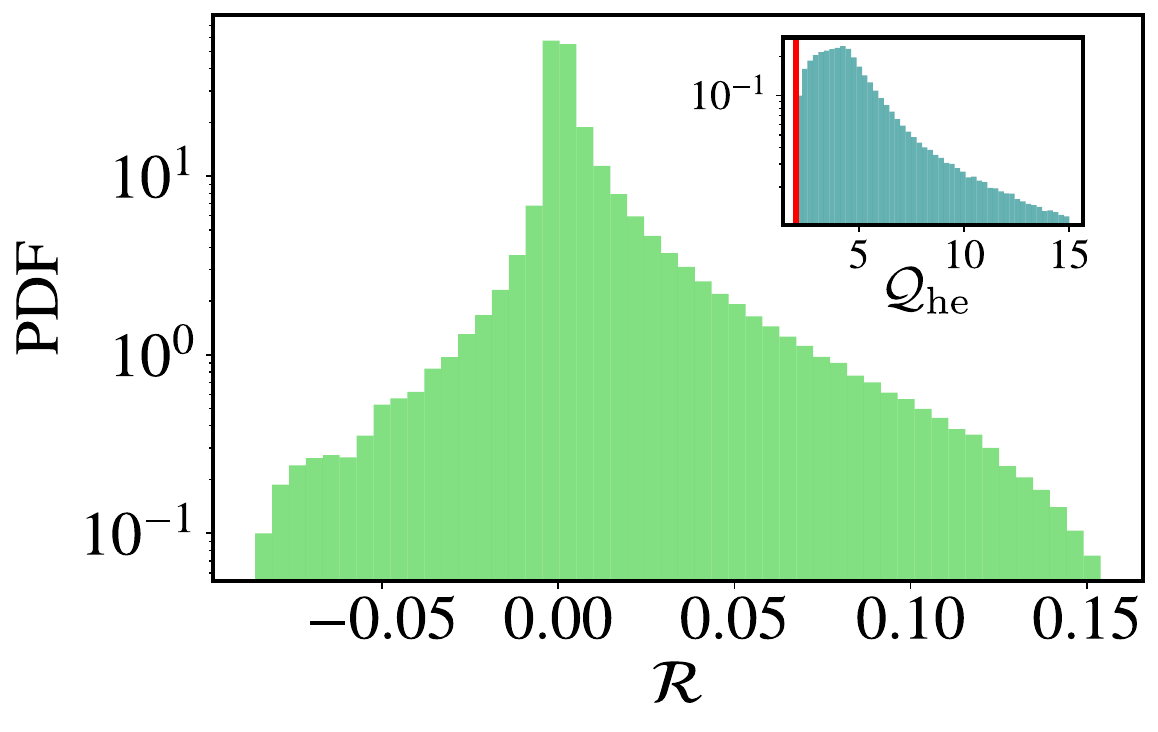}
  \caption{Histogram for the NIC machine of sampled values of the fluctuation ratio $\mathcal{R}$ and TUR ratio $\mathcal{Q}_{\rm he}$ (inset plot). The negative values of $\mathcal{R}$ indicate configurations for which the classical equivalent machine has reduced fluctuations with respect to the original quantum one. The values corresponds to an exploration of the region in the parameters space: $\beta_{\mathrm{h}}/\beta_{\mathrm{c}} \in [0,1]$, $\epsilon_{1} \in [0.1,4.9]$, $\gamma_{\mathrm{h}}^{(\mathrm{a}/\mathrm{b})} \in [10^{-5}, 10^{-2}]$. Fixed parameters are  $\beta_{\mathrm{c}} = 1$, $\beta_{\mathrm{w}} \rightarrow 0$, $\epsilon_{2} = 5$, and $\gamma_{\mathrm{c}}^{(\mathrm{a}/\mathrm{b})} = 10^{-3}$ (energies are given in $k_{\rm B} T_{\rm c} = 1$ units). }
  \label{fig: fridges results 2}
\end{figure}

We obtain analytically the fluctuation ratio $\mathcal{R}$ in Eq.~\eqref{eq:R} for the NIC machine and its classical equivalent, taking again $J_\mathrm{out} = \dot{W}$. By evaluating the expression for $10^6$ different parameter configurations within the region where the classical equivalent machine can be defined, we show in Fig.~\ref{fig: fridges results 2} the distribution of $\mathcal{R}$ together with the corresponding distribution of the TUR ratio $\mathcal{Q}_\mathrm{he}$ values (see inset). Our results show that $\mathcal{R} \geq 0$ for some configurations, pointing to quantum-thermodynamic advantages in a similar way than in the two previous examples. However, here we also get configurations leading to $\mathcal{R} < 0$. For that configurations, the quantum-coherent machine becomes more noisy than its classical equivalent, hence leading to a detrimental role for the coherence. Indeed, for some configurations the NIC machine reaches noise levels up to $1\%$ higher than its classical counterpart. Moreover, by looking at the inset we observe that, as expected, the TUR is not violated for all choices of parameters ($\mathcal{Q}_\mathrm{he} \geq 2$ always) in accordance with previous studies about TUR violations in similar models~\cite{Segal21}.

Therefore, it becomes clear that the NIC machine is able to overcome the classical equivalent machine in some configurations which are not witnessed by TUR violations as in the case of Hamiltonian-induced coherence between degenerate levels; but not in all configurations. Contrary to the case of Hamiltonian-induced coherence, noise-induced one does not necessarily have a beneficial impact on the thermodynamic performance of the NIC machine, whose operation in terms of power, efficiency, and reliability can be surpassed in some regimes by a purely classical equivalent machine using the same set of resources. 

The region of parameters in which a quantum-thermodynamic advantage over the classical equivalent is verified can be obtained analytically and is given by:
\begin{equation}\label{eq:conditions_amplification}
\dfrac{\gamma_{\mathrm{h}}^{\alpha}}{\gamma_{\mathrm{h}}^{\beta}} \geq \dfrac{2 + 3 (\bar{n}_{\mathrm{h}}+\bar{n}_{\mathrm{c}}) + 4 \bar{n}_{\mathrm{h}}\bar{n}_{}\mathrm{c}}{\bar{n}_{\mathrm{h}}-\bar{n}_{\mathrm{c}}},\;\;\;\gamma_{\mathrm{h}}^{\beta}\neq 0. 
\end{equation}
Detrimental effects from coherence are obtained for regions where the above inequality is inverted. In particular, in the symmetric case $\gamma_{\rm c}^{\rm a} = \gamma_{\rm c}^{\rm b}$ and $\gamma_{\rm h}^{\rm a} = \gamma_{\rm h}^{\rm b}$ we will have $\gamma_{\mathrm{h}}^{\beta} = 0$ and the system effectively becomes a three-level system without coherence in the steady state. 
In this limit, the quantum system and its classical analog become indistinguishable. 
These conclusions are to be compared with previous results in the literature for quantum enhancements (also in the symmetric case) for same or similar NIC models, see e.g. Refs.~\cite{Scully11,Dorfman13,Gelbwaser15,Um22}.

\section{Conclusions}
\label{sec:conclusions}

We have characterized the thermodynamic impact of Hamiltonian and noise-induced coherence in the performance of quantum thermal machines operating in nonequilibrium steady states. Machines displaying Hamiltonian coherence in their steady states lead to quantum-thermodynamic advantages in terms of their trade-off between power, efficiency, and stability (Result 1). However, this is not always the case for machines that display noise-induced coherence (Result 2). These results, obtained through direct comparison of generic quantum thermal machines with axiomatically constructed classical equivalent models, ensure the presence of quantum-enhanced performance for models of Hamiltonian-induced coherence in out-of-equilibrium situations. However, they also imply that several previous claims of thermodynamic advantages in machines with noise-induced coherence may be spurious and suggest that they should be reassessed on a case-by-case basis, by determining the set of parameters in which quantum advantages may really exist (if any). Indeed, even if the dynamics of these machines may exhibit purely quantum features, it turns out that a classical Markovian machine using the same set of resources (energy gaps and bath temperatures) might be systematically constructed that performs just as well or better than the original machine.

The classical equivalent model employed here allowed us to compare the current fluctuations between a quantum device and a classical (incoherent) counterpart that outputs the same average currents while employing identical thermodynamic resources. This notion of a classical equivalent machine builds on the idea of emulability discussed in Ref.~\cite{Onam19}, and is accompanied by a general methodology for its derivation in generic cases. The classical equivalent can be constructed for virtually any quantum steady-state machine working in the weak-coupling regime, and under weak-driving conditions, namely, when the driving can be considered a perturbation of the (bare) machine Hamiltonian. Extensions of this method to the case of strong couplings or strong periodic driving are an interesting direction for future research, which may allow the addressing of quantum thermodynamic advantages in a larger class of quantum devices.

Using the three-level amplifier and the three-qubit-fridge as main examples of a quantum thermal machine displaying Hamiltonian coherence (the first energetic coherence and the second degenerate one), we have illustrated situations where coherence is always beneficial. In these cases, the performance matches or exceeds the performance of its classical equivalent counterpart in all possible configurations, with improvements on the machine stability that become greater far from equilibrium. In the case of the three-level amplifier, the parameter regions where these improvements are maximal coincide indeed with regimes where the thermal machine breaks the TUR bound. However, we also observe wide regions showing thermodynamic improvements that are not witnessed by TUR violations. This is indeed entirely the case for the three-qubit fridge, where the TUR by itself is not able to spot thermodynamic enhancements. Importantly, our results reveal that to observe a genuine quantum-thermodynamic advantage, it is necessary (and sufficient) to compare the fluctuations in the currents (at least at the level of the variance), in contrast to previous assessments comparing the average currents and based on more limited notions of classical equivalents (leading to a less stringent comparison)~\cite{Uzdin15}. Moreover, while in the illustrative examples we considered resonant driving for simplicity, Result~1 ensures that our findings remain robust in the presence of detuning within the range $|\Delta_{\rm d}|< (\gamma_{\rm c} n_{\rm c} + \gamma_{\rm h} n_{\rm h})/2$.

In this work we focused on continuous thermal machines operating in steady-state conditions under weak coupling conditions and weak driving. Future research could focus on exploring other classes of machines, by analyzing the possibility of defining classical equivalents in regimes where the weak driving approximation does not hold (by using, e.g., Floquet formalism~\cite{Cuetara2015,Gelbwaser15}), by studying the effects of strong coupling~\cite{Segal23} and non-Markovian evolution~\cite{Restrepo2018}, or by including the presence of quantum resources in the reservoirs such as squeezing~\cite{Correa14} or coherence~\cite{Hammam24}. This program would allow us to investigate whether genuine quantum thermodynamic advantages can be identified and certified in broader scenarios.

\begin{acknowledgments}
We thank Roberta Zambrini, Mark T. Mitchison, and Camille L. Latune for comments and interesting discussions. We wish to acknowledge support from the CoQuSy project PID2022-140506NB-C21 and the Mar\'ia de Maeztu project CEX2021-001164-M for Units of Excellence, funded by MICIU/AEI/10.13039/501100011033/FEDER, UE. GM acknowledges the Ram\'on y Cajal program RYC2021-031121-I funded by MICIU/AEI/10.13039/501100011033 and European Union NextGenerationEU/PRTR. JAAM acknowledges Conselleria d'Educaci\'o, Universitat i Recerca of the Balearic Islands (Grant FPI$\_058\_2022$).
\end{acknowledgments}

\bibliographystyle{quantum}
\bibliography{referencias.bib}

\onecolumn\newpage
\appendix

\section{Quantum thermal machines models}
\label{sec:modelsmethods}

Here we provide a detailed description of the quantum thermal machine models analyzed as representative examples in this work. Specifically, we examine three key cases: a driven three-level amplifier, an autonomous three-qubit absorption refrigerator, and a noise-induced coherence machine.

\subsection{Coherent three-level amplifier} \label{sec:model-amplifier}

The three-level maser or amplifier, as initially introduced by Scovil and Schulz-DuBois in Ref.~\cite{Scovil}, is one of the simplest models of a thermal machine, capable of performing either as a heat engine or a refrigerator, depending on the configuration of system parameters~\cite{Geusic67,Palao01}.
The characteristics and performance of this model have been largely studied~\cite{Geva94,Palao01,Boukobza07,Kosloff14,Li17,Dorfman18, Binder18, Vardiner20,Kalaee21} owing to its simplicity and versatile functionality, and a first experimental implementation has been reported in Ref.~\cite{Klatzow19}. It stands as a main example of a thermal machine where coherence among non degenerate energy levels is supported in the steady state, suggesting the appearance of regimes where quantum-enhanced thermodynamic operation can be achieved~\cite{Uzdin15} (for a critical view of such quantum-enhanced performance, see e.g. Ref.~\cite{Onam19}). 

The machine system contains three discrete energy levels and an external driving field acting on its lower transition, as depicted in Fig.~\ref{fig:diagrams}a. The Hamiltonian of the system can be written in this case as: 
\begin{equation}\label{eq: Hamiltonian Amplifier}
    H = \epsilon_{1} \ket{1}\bra{1} + \epsilon_{2} \ket{2}\bra{2} + V(t),
\end{equation}
with bare Hamiltonian $H_0 := \epsilon_{1} \ket{1}\bra{1} + \epsilon_{2} \ket{2}\bra{2}$ time dependent driving Hamiltonian $V(t) = g (e^{i\omega_\mathrm{d} t} \ket{0} \bra{1} + \text{h.c})$ which we consider for simplicity resonant with the first energy gap, i.e. $\omega_{d} \equiv \epsilon_1$, and where $g$ is the external driving field strength. In the absence of thermal baths, this driving field generates Rabi-like oscillations within the first two levels of the machine, $\ket{0}$ and $\ket{1}$, at frequency $g$. 

The two remaining transitions of the three-level system are weakly coupled to two thermal baths at different inverse temperatures, denoted as cold and hot ($\beta_\mathrm{c} \geq \beta_\mathrm{h}$), typically modeled as bosonic reservoirs. They lead to four incoherent jumps described by Lindblad operators:
\begin{equation}\label{eq: lindblad operator Amplifier}
    \begin{split}
        L_{\downarrow}^{(\mathrm{c})} &= \sqrt{\gamma_{21}} \ket{1}\bra{2};~~ L_{\uparrow}^{(\mathrm{c})} = \sqrt{\gamma_{12}} \ket{2}\bra{1}, \\
        L_{\downarrow}^{(\mathrm{h})} &= \sqrt{\gamma_{20}} \ket{0}\bra{2};~~  L_{\uparrow}^{(\mathrm{h})} = \sqrt{\gamma_{02}} \ket{2}\bra{0},
    \end{split}
\end{equation}
with rates $\gamma_{21} = \gamma_\mathrm{c} (\bar{n}_\mathrm{c} + 1)$ and $\gamma_{12} = \gamma_\mathrm{c} \bar{n}_\mathrm{c}$ associated respectively to the emission and absorption of energy quanta $\Delta \epsilon_\mathrm{c} = \epsilon_2 - \epsilon_1$ into the cold reservoir at $\beta_\mathrm{c}$. Similarly $\gamma_{20} = \gamma_\mathrm{h} (\bar{n}_\mathrm{h} + 1)$ and $\gamma_{02} = \gamma_\mathrm{h} \bar{n}_\mathrm{h}$ stand for emission and absorption of energy quanta $\Delta \epsilon_\mathrm{h} = \epsilon_2$ into the hot reservoir at temperature $\beta_\mathrm{h}$. Here, $\gamma_r$ denotes the spontaneous emission rate for bath $r = \mathrm{c, h}$ and $\bar{n}_r = (e^{\beta_r \Delta \epsilon_r} -1)^{-1}$ is the average number of thermal photons with frequency $\Delta \epsilon_r$ in the baths.

As commented before, since we are treating the interaction Hamiltonian as a perturbation, our analysis will be limited to the weak driving regime. In this regime, it is possible to extract the time dependence of the system's Hamiltonian by moving to a rotating frame (interaction picture with respect to $H_0$). This transformation leads to the following form of the master equation~\eqref{quantum_master_equation}:
\begin{equation}
    \dfrac{d\rho^{\prime}}{dt} = -i \left [ V_I, \rho^{\prime}\right] + \sum_{r= \mathrm{c, h}} \sum_{k = \downarrow, \uparrow} \mathcal{D}^{(r)}_{k}[\rho^\prime],
\end{equation}
where  $\rho^{\prime} = e^{i H_{0} t}\rho e^{- i H_{0} t}$ and the driving Hamiltonian in the interaction picture appearing in the Lindblad equation has the simpler form $V_I = g \left ( \ket{0} \bra{1} +  \ket{1} \bra{0}\right)$. 

\subsection{Three-qubit autonomous refrigerator} \label{sec:model-fridge}

This fridge model was first presented in Ref.~\cite{Linden10} and consists of one of the smallest thermal machine models using a multipartite system (see Fig. \ref{fig:diagrams}b). It pertains to the class of autonomous quantum refrigerators~\cite{Skrzypczyk11,Brunner12,Silva16}, also called quantum absorption refrigerators~{\cite{Levy12,Correa13,Correa14,Correa14b,Mitchison15,Seah18}. It consists of three qubits with different energy gaps, each of them locally coupled to a corresponding thermal bath at a different temperature. A weak three-body energy-preserving Hamiltonian interaction among the qubits allows the generation of heat currents between the three reservoirs that ultimately power thermodynamic tasks such as heat pumping or refrigeration~\cite{Brunner12}. Different platforms for the actual implementation of this or closely related models in the lab, have been proposed using quantum dots~\cite{Venturelli13}, optical systems~\cite{Mitchison16}, QED architectures~\cite{Hofer16} and trapped ions~\cite{Maslennikov19}.  In this multipartite setup, genuine quantum features such as quantum entanglement~\cite{Brunner14} and discord~\cite{Correa13} have been analyzed, suggesting the possibility of boosting the fridge performance by quantum correlations in some operational regimes~\cite{Brunner14}.

The Hamiltonian of the three-qubit working substance here reads:
\begin{equation}\label{eq: Hamiltonian QAR}
    H = H_{1} + H_{2} + H_{3} + V, 
\end{equation}
where we identify $H_0 = \sum_i H_i$ with $H_{i} = \epsilon_{i} \ket{1}\bra{1}_{i}$ the (bare) Hamiltonians of each individual qubit and $V = g \left ( \ket{101}\bra{010} + \ket{010}\bra{101}  \right)$ is a three-body interaction Hamiltonian allowing the qubits to exchange energy (notice the use of the simplified notation $\ket{101} \equiv \ket{1}_1\ket{0}_2\ket{1}_3$, etc.). Here above $g \ll \epsilon_i$, ensures that the interaction can be treated as a perturbation to the bare three-qubit Hamitlonian $H_0$. 

Importantly, by assuming the resonance condition $\epsilon_3 = \epsilon_2- \epsilon_1$, the three-qubit interaction verifies strict energy preservation between the qubits, that is, $[V, H_1 + H_2 + H_3] = 0$, ensuring that the energy exchanges among the fridge qubits occur without the need of any extra source of energy or control, i.e. preserving the autonomy of the model. At difference from other autonomous fridges, such as single-qutrit fridges~\cite{Palao01}, this model exhibits steady-state coherence between degenerate energy levels $\ket{101}$ and $\ket{010}$ due to the presence of interaction $V$~\cite{Linden10}, which ultimately leads to entanglement among different partitions involving the qubits~\cite{Brunner14}. 

In this case, all transitions are mediated by the reservoirs, with either cold, mild, or hot temperatures ($\beta_\mathrm{c} \geq \beta_\mathrm{m} \geq \beta_\mathrm{h}$). Since each qubit $i$ is locally coupled only to a single bath at inverse temperature $\beta_i$, and the interaction $V$ is weak, the master equation~\eqref{quantum_master_equation} adopts a local form~\cite{Trushechkin16,Hofer17,Chiara18}, with six incoherent jumps described by Lindblad operators promoting local jumps in each qubit:
\begin{equation}\label{eq: lindblad operators QAR}
    \begin{split}
        L_{\downarrow}^{(\text{c})} &= \sqrt{\gamma_{10}^\text{c}} \ket{0}\bra{1}_1 \otimes \mathds{1}_2 \otimes \mathds{1}_3, \\
        L_{\downarrow}^{(\text{m})} &= \sqrt{\gamma_{10}^\text{m}} \mathds{1}_1 \otimes \ket{0}\bra{1}_2 \otimes \mathds{1}_3, \\
        L_{\downarrow}^{(\text{h})} &= \sqrt{\gamma_{10}^\text{h}} \mathds{1}_1 \otimes \mathds{1}_2 \otimes \ket{0}\bra{1}_3, \\
    \end{split}
\end{equation}
together with the opposite jumps, $L_{\uparrow}^{(r)} =e^{-\beta_{r}\Delta \epsilon_{r}/2} L_{\downarrow}^{(r)\dagger}$, for $r = \text{c}, \text{m}, \text{h}$. The rates $\gamma_{10}^{\text{c}} = \gamma_\mathrm{c} (\bar{n}_\mathrm{c} + 1)$ and $\gamma_{01}^{\text{c}} = \gamma_\mathrm{c} \bar{n}_\mathrm{c}$ are associated, respectively, to the emission and absorption of energy quanta $\Delta \epsilon_\mathrm{c} = \epsilon_1$ into the cold reservoir at $\beta_\mathrm{c}$. Similarly, we have $\gamma_{10}^{\text{m}} = \gamma_\mathrm{m} (\bar{n}_\mathrm{m} + 1)$ and $\gamma_{01} = \gamma_\mathrm{m} \bar{n}_\mathrm{m}$, for the emission and absorption of energy quanta $\Delta \epsilon_\mathrm{m} = \epsilon_2$ into the medium reservoir at temperature $\beta_\mathrm{m}$, as well as $\gamma_{10}^{\text{h}} = \gamma_\mathrm{h} (\bar{n}_\mathrm{h} + 1)$ and $\gamma_{01}^{\text{h}} = \gamma_\mathrm{h} \bar{n}_\mathrm{h}$ for the emission and absorption of energy quanta $\Delta \epsilon_\mathrm{h} = \epsilon_3$ into the hot reservoir at temperature $\beta_\mathrm{h}$.

\subsection{Noise-induced-coherence machine} \label{sec:model-noise}

A final set of continuous thermal machine models showing quantum effects, which we collectively dub noise-induced-coherence (NIC) machines, were first presented in a series of papers by Scully {\it et al.} \cite{Scully10,Scully11,Dorfman13}. In these machines, degenerate levels in the energy spectrum are combined with a collective action of the baths on the system transitions to generate coherence in the steady state~\cite{Kozlov06,Tscherbul14}. The operation and performance of these kind of machines has been extensively investigated within the context of quantum thermal machines~\cite{Harbola12,Harbola13,Gelbwaser15,Killoran15,Su16, Dorfman18,Holubec18,Manzano19,Latune19,Segal21,Um22}, pointing to output power enhancements for adequate operational regimes.

A notable example of this coherence is found in a 4-level absorption refrigerator that can also work as a heat engine, whose thermodynamics has been examined in previous studies~\cite{Holubec18, Holubec19, Segal21}. In this system, two of the levels possess the same energy, as depicted in Fig.~\ref{fig:diagrams}c. These levels are subjected to the influence of two distinct thermal baths whose action results in the emergence of degenerate coherence that persist in the steady state. The system's Hamiltonian in this case is as follows
\begin{equation}\label{eq: Hamiltonian NIC}
    H =\epsilon_{1} \ket{1}\bra{1} + \epsilon_{2} \left( \ket{2\text{a}}\bra{2\text{a}} + \ket{2\text{b}}\bra{2\text{b}}\right) = H_0.  
\end{equation}
In this case all the transitions are mediated by the reservoirs at cold, hot and ``work" temperatures ($\beta_\mathrm{c} \geq \beta_\mathrm{h} > \beta_\mathrm{w}\rightarrow 0$). To represent these transitions we will have six different incoherent jumps, two of them consisting of individual jumps $L_{\downarrow}^{(\text{w})} = \sqrt{\gamma_{10}}\ket{0}\bra{1}$ and $L_{\uparrow}^{(\text{w})} = \sqrt{\gamma_{01}}\ket{1}\bra{0}$, and the other four refer to collective transitions: 
\begin{equation}\label{eq:lindblad operators NIC}
    \begin{split}
        L_{\downarrow}^{(\text{h})} &  =\sqrt{\gamma_{\text{a}1}}\ket{1}\bra{2\text{a}} + \sqrt{\gamma_{\text{b}1}}\ket{1}\bra{2\text{b}}, \\
        L_{\uparrow}^{(\text{h})} &= \sqrt{\gamma_{1\text{a}}}\ket{2\text{a}}\bra{1} + \sqrt{\gamma_{1\text{b}}}\ket{2\text{b}}\bra{1}  ,\\
        L_{\downarrow}^{(\text{c})} &= \sqrt{\gamma_{\text{a}0}}\ket{0}\bra{2\text{a}} + \sqrt{\gamma_{\text{b}0}}\ket{0}\bra{2\text{b}}  ,\\
        L_{\uparrow}^{(\text{c})} &= \sqrt{\gamma_{0\text{a}}}\ket{2\text{a}}\bra{0} + \sqrt{\gamma_{0\text{b}}}\ket{2\text{b}}\bra{0},  
    \end{split}
\end{equation}
with rates $\gamma_{i1} = \gamma_\mathrm{h}^{i} (\bar{n}_\mathrm{h} + 1)$ and $\gamma_{1i} = \gamma_\mathrm{h}^{i} \bar{n}_\mathrm{h}$ for $i = \text{a}, \text{b}$, associated respectively to the emission and absorption of energy quanta $\Delta \epsilon_\mathrm{h} = \epsilon_2 - \epsilon_1$ into the hot reservoir at $\beta_\mathrm{h}$, and similarly $\gamma_{i0} = \gamma_\mathrm{c}^{i} (\bar{n}_\mathrm{c} + 1)$ and $\gamma_{0i} = \gamma_\mathrm{c}^{i} \bar{n}_\mathrm{c}$ for $i = \text{a}, \text{b}$, is associated with emission and absorption of energy quanta $\Delta \epsilon_\mathrm{c} = \epsilon_2$ into the cold reservoir at temperature $\beta_\mathrm{c}$. Finally, since the ``work" bath is at an infinite temperature i.e $\beta_{\text{w}} \rightarrow 0$ the rates associated to it satisfy $\gamma_{10} = \gamma_{01}$, corresponding in this case for emission and absorption of a quanta $\Delta \epsilon_{\text{w}} = \epsilon_1$ from the work source.

\section{Thermodynamic performance}
\label{sec:performance}

We are interested in the performance of the quantum thermal machine models presented in the previous section, when operating in nonequilibrium steady-state conditions~\cite{Kosloff14,Binder18}. By performance, we refer not only to the size of the output current generated by the machine operation (output power in the case of heat engines or cooling power for the case of refrigerators) and the thermodynamic efficiency of the machine (ratio of useful output to source input), but also to the size of the fluctuations in the output current, which can be viewed as a measure of the ``quality'' of that output in stochastic machines~\cite{Pietzonka18}.

Under steady-state conditions, $\mathcal{L}(\pi) = 0$, the average output power generated by the machine on the external drive and the average heat current absorbed from reservoir $r$, are given, respectively, by standard definitions~\cite{Alicki79}: 
\begin{align} \label{heat_i}
\langle \dot{W} \rangle &:= - \Tr[\dot{H}(t) \pi(t)] = - \Tr[\dot{V}(t) \pi(t)], \\
\langle \dot{Q}_r \rangle &:= \sum_k \Tr[H \mathcal{D}_k^{(r)}[\pi(t)]] \simeq \sum_k \Tr[H_0 \mathcal{D}_k^{(r)}[\pi(t)]], \nonumber
\end{align}
where we recall that $\pi(t)$ may acquire a periodic time-dependence (in Schr\"odinger picture) due to the presence of non-diagonal elements (coherences) in the steady-state density operator. We also emphasize that for weak perturbations $V$ as the ones considered here, in order to be consistent with the microscopic derivation of the master equation~\cite{Trushechkin16}, only the bare Hamiltonian $H_0$ should enter in the heat currents, which ensures consistency with the second law~\cite{Hewgill21,Manzano18,Manzano22} (see also App. \ref{appendix: general expression heat}). The first law takes the form $\langle \dot{W} \rangle = \sum_r \langle \dot{Q}_r \rangle$, imposing that any output power is sustained by input heat currents from the baths. Explicit expressions of the heat currents $\langle \dot{Q}_r \rangle$ valid for generic machines (with or without degeneracy) are given in Appendix \ref{appendix: general expression heat}.

As a consequence of Markovianity, the second law in the setup is manifested through the non-negativity of the entropy production rate:
\begin{equation}\label{ec ent prod}
    \langle\dot{S}_\mathrm{tot}\rangle = -\sum_{r}\beta_{r}\langle\dot{Q}_{r}\rangle  \geq 0,
\end{equation}
which characterizes the irreversibility of the machine operation in its nonequilibrium steady-state~\cite{Manzano18,Landi21}. Here, it is also worth remarking that, whenever the temperature of some of the reservoirs $r$ approaches infinity, $\beta_r \rightarrow 0$, the associated energy current does not contribute to the entropy production, and hence it should be considered as (incoherent) work rather than heat, see also Refs.~\cite{Levy12,Strasberg17}. In that case, the output power associated to such a work reservoir reads $\langle \dot{W} \rangle = -\sum_k \Tr[H_0 \mathcal{D}_k^{(r)}[\pi(t)]]$.   

The efficiency of thermal machines can be defined from the ratio of the average output useful current to the average input resource one, as determined by the operational mode of the machine: 
\begin{equation}
    \eta := \frac{\langle {J}_\mathrm{out} \rangle}{ \langle {J}_\mathrm{in} \rangle},
\end{equation}
where in the case of heat engine operation ${J}_\mathrm{out} = \langle \dot{W} \rangle$ and ${J}_\mathrm{in} = \langle \dot{Q}_\mathrm{h} \rangle$, while for refrigeration the efficiency (coefficient of performance) is given from ${J}_\mathrm{out} = \langle \dot{Q}_\mathrm{c} \rangle$ and either ${J}_\mathrm{in} = - \langle \dot{W} \rangle$ for power-driven refrigerators (as the models in Fig.~\ref{fig:diagrams}a and \ref{fig:diagrams}c), or ${J}_\mathrm{in} = - \langle \dot{Q}_\mathrm{h} \rangle$ for absorption refrigerators (as the one in Fig.\ref{fig:diagrams}b). For extensions of efficiency to multiple inputs and  outputs see, e.g. Refs.~\cite{Potts20,Lopez23,Hammam24}.

By combining the first and second laws in the setup, we recover Carnot bound for the efficiency of heat engines $\eta \leq \eta_\mathrm{C} := 1 - {\beta_\mathrm{h}/\beta_\mathrm{c}}$, as well as the corresponding (Carnot) bounds for power-driven fridges $\eta \leq \eta_\mathrm{fr} :=  \beta_\mathrm{h}/(\beta_\mathrm{c} - \beta_\mathrm{h})$ and absorption refrigerators, $\eta \leq \eta_\mathrm{abs} := (\beta_\mathrm{m} - \beta_\mathrm{h})/(\beta_\mathrm{c} - \beta_\mathrm{m})$, achieved in the limit of zero entropy production, where all energy currents vanish~\cite{Kosloff14,Binder18,Skrzypczyk11}. 

The maximum efficiency (zero power) equilibrium point separates the different modes of operation of the machines, where heat currents and output power change sign. In local dissipation models, due to the absence of heat leaks, the average steady-state currents can be rewritten as:
\begin{equation}\label{eq:current stationarity}
\langle \dot{Q}_{r} \rangle = \Delta\epsilon_{r} \langle \dot{N}_r \rangle, 
\end{equation}
where $\langle \dot{N}_r \rangle$ is the probability current associated to the reservoir $r$ (flux of quanta). This expression comes from the baths being associated with a single energy gap $\Delta\epsilon_r$ (the eventual action of a bath over other energy gaps may be considered as produced by an independent bath at the same temperature). In the steady state, if the system exchanges excitations with all baths at the same rate (tight coupling), we have $\forall i,j$ that $\langle \dot{N}_i \rangle =\langle \dot{N}_j \rangle := \langle \dot{N} \rangle$. In addition, the first law constrains the average value of the work to satisfy $\langle \dot{W} \rangle = \sum_{r} \Delta \epsilon_{r} \langle \dot{N} \rangle = \omega_d \langle \dot{N} \rangle$. The ratio between the different steady-state currents hence verifies
\begin{equation}\label{eq: stationary relation}
   \frac{|\langle \dot{W} \rangle|}{|\langle \dot{Q}_j\rangle|} = \frac{\omega_\mathrm{d}}{\Delta \epsilon_j} ~~~;~~~ \frac{|\langle \dot{Q}_i \rangle|}{|\langle \dot{Q}_j\rangle|} = \frac{\Delta \epsilon_i}{\Delta \epsilon_j},
\end{equation}
which is verified in the three illustrative examples. By combining the above proportionality relations with the efficiency bounds, we can construct a table for the operational modes of each model (see Table~\ref{tab:1}).
\begin{table}[t]
    \centering
    \begin{tabular}{| r | l | c |} \hline
     & ~Heat engine/pump~ & ~Refrigerator~ \\ \hline
    Three-level amplifier & $ ~~~~~\omega_{\mathrm{d}}/\epsilon_2 \leq \eta_\mathrm{C}$  & $\eta_\mathrm{C} \leq \omega_{\mathrm{d}}/\epsilon_2 \leq 1$ \\
    Autonomous fridge~ & $~~ \eta_\mathrm{abs}\leq \epsilon_1/\epsilon_3 \leq 1$ & $\epsilon_1/\epsilon_3 \leq \eta_\mathrm{abs}$ \\
    NIC machine~~~~~ & $~~~~~~ \epsilon_1/\epsilon_2 \leq \eta_\mathrm{C}$ & $~\eta_\mathrm{C} \leq \epsilon_1/\epsilon_2 \leq 1~$ \\ \hline
    \end{tabular}
    \caption{Parameter relations leading to the main modes of operation of the three thermal machines examined here. In the case of the autonomous (absorption) refrigerator the heat engine regime is replaced by heat pumping.}
    \label{tab:1}
\end{table}

In addition to minimizing entropy production and maximizing power, low fluctuations (resulting in higher precision) in energy flows are also desirable for the performance of thermal machines. In classical systems these three quantities are not independent, but their trade-off is quantified by the TUR:
\begin{equation}\label{TUR}
    \mathcal{Q} = \langle \dot{S}_{\text{tot}} \rangle \dfrac{\text{Var}[J_\mathrm{out}]}{\langle J_\mathrm{out} \rangle^{2}} \geq 2,
\end{equation}
where $\langle J_\mathrm{out} \rangle$ and $\text{Var}[J_\mathrm{out}]$ denote the mean and variance of the useful output current i.e $J_\mathrm{out} = \dot{W}, \dot{Q}_\mathrm{c}$ for work production and refrigerator regimes respectively. Initially proposed in the context of bio-molecular processes~\cite{Barato15}, the TUR was then formally established in stochastic thermodynamics~\cite{Todd16,Horowitz20}, and subsequently applied to classical steady-state heat  engines~\cite{Pietzonka18}, for which the TUR ratio $\mathcal{Q}$ in Eq.~\eqref{TUR} can be rewritten in terms of the output power and efficiency as:
\begin{equation} \label{eq:Qengine}
\mathcal{Q}_\mathrm{he} = \beta_\mathrm{c}  \dfrac{\text{Var}[\dot{W}]}{\langle \dot{W} \rangle} \left( \frac{\eta_\mathrm{C} - \eta}{\eta} \right),
\end{equation}
where we identified $J_\mathrm{out} = \dot{W}$ and $\langle \dot{S}_\mathrm{tot} \rangle = -\beta_\mathrm{h} \langle \dot{Q}_\mathrm{h} \rangle - \beta_\mathrm{c} \langle \dot{Q}_\mathrm{c} \rangle = \beta_\mathrm{c} (\eta_\mathrm{C}/\eta - 1) \langle \dot{W} \rangle$. Analogously by taking 
$J_\mathrm{out} = \dot{Q}_\mathrm{c}$ and rewriting the expression for the entropy production rate, we obtain the corresponding TUR ratio for power-driven refrigerators:
\begin{equation}
 \mathcal{Q}_\mathrm{fr} =  (\beta_\mathrm{c} - \beta_\mathrm{h} )\dfrac{\text{Var}[\dot{Q}_\mathrm{c}]}{\langle \dot{Q}_\mathrm{c} \rangle} \left( \frac{\eta_\mathrm{fr} - \eta}{\eta} \right).   
\end{equation}
Finally, for the case of absorption refrigerators we obtain the TUR ratio:
\begin{equation} \label{eq:Qabs}
 \mathcal{Q}_\mathrm{abs} =  (\beta_\mathrm{c} - \beta_\mathrm{m} )\dfrac{\text{Var}[\dot{Q}_\mathrm{c}]}{\langle \dot{Q}_\mathrm{c} \rangle} \left( \frac{\eta_\mathrm{abs} - \eta}{\eta} \right),
\end{equation}
where in this case we identified $J_\mathrm{out}= \dot{Q}_\mathrm{c}$ and the entropy production rate reads $\langle \dot{S}_\mathrm{tot} \rangle = -\beta_\mathrm{h} \langle \dot{Q}_\mathrm{h} \rangle - \beta_\mathrm{c} \langle \dot{Q}_\mathrm{c} \rangle - \beta_\mathrm{m} \langle \dot{Q}_\mathrm{m} \rangle = (\beta_\mathrm{c} - \beta_\mathrm{m}) (\eta_\mathrm{abs}/\eta - 1) \langle \dot{Q}_\mathrm{c} \rangle$. Throughout this paper, to evaluate fluctuations in the output currents, i.e. the variances $\text{Var}[\dot{W}]$ and $\text{Var}[\dot{Q}_\mathrm{c}]$, we employ the full-counting statistics (FCS) formalism \cite{Esposito09, Bruderer14}.

In any of the three cases, the TUR implies that beyond a certain threshold, a classical Markovian engine can only enhance its precision in the output (cooling) power at the cost of either reducing the output itself or reducing the energy efficiency, so that the above ratio remains bounded by $2$, i.e. $\mathcal{Q} \geq 2$. However, some models of quantum thermal machines have been shown to produce violations of the TUR, that is, they verify $\mathcal{Q} < 2$ (see e.g. Refs.~\cite{Ptaszy18,Liu19,Kalaee21,Mitchison21,Souza22,Manzano23}). Such violations act as a witness indicating an enhanced trade-off between power, precision, and efficiency that arises in certain parameter regimes. 
Nevertheless, such quantum-thermodynamic advantages may arise even if the TUR is not violated, since classical machines may not saturate the TUR in relevant operational regimes. Hence in order to provide a fair and accurate assessment of quantum-thermodynamic advantages in thermal machines, comparison to classical models becomes necessary.

\section{General expression of current first moments with local dissipation}\label{appendix: general expression heat}

In this Appendix we derive a general expression for the average heat currents including the cases in which coherence may be generated either by Hamiltonian or noise-induced sources. We start by expanding the expression for the standard definition for the heat currents [Eq. ~\eqref{heat_i}], that is: 
\begin{align}\label{eq:heat_derivation}
    &\langle \dot{Q}_r \rangle = \sum_k \Tr[(H_0 + V(t)) \mathcal{D}_k^{(r)}[\pi(t)]] \\&\simeq \sum_k \Tr[H_0 \mathcal{D}_k^{(r)}[\pi(t)]] = \sum_k \Tr[\mathcal{D}_k^{\dagger (r)}[H_0]\pi(t)] \nonumber \\
    &= \sum_{k}\sum_{m,l} \bra{m} L_{k}^{(r)\dagger}H_{0}L_{k}^{(r)} - \dfrac{1}{2} \{ L_{k}^{(r)\dagger}L_{k}^{(r)}, H_{0}\}\ket{l} \pi_{lm} \nonumber
\end{align}
where the third-order terms $\Tr[V(t)\mathcal{D}_k^{(r)}[\pi(t)]] \sim g \gamma_0^{(r)}$, with $|V|\sim g$ and $\gamma_k^{(r)}$ being the spontaneous emission rate in reservoir $r$ and transition $k$, have been neglected in the second line in accordance to the weak coupling and weak driving approximations. In the second line above, we have applied the cyclic property of the trace to obtain the adjoint dissipator $\mathcal{D}_k^{\dagger (r)}[\cdot] := L_{k}^{(r)\dagger} \cdot L_{k}^{(r)} - \dfrac{1}{2} \{ L_{k}^{(r)\dagger}L_{k}^{(r)}, \cdot \}$. 

Using the explicit form of the Lindblad operators in Eq.~\eqref{eq:Lindblad-ops} and the bare Hamiltonian, $H_0 = \sum_i \epsilon_i \ket{i}\bra{i}$, we can obtain each term in the sum of the last expression:  
\begin{equation}\label{eq: terms_heat}
\begin{split}
&L_{k}^{(r)\dagger}H_{0}L_{k}^{(r)} = \sum_{i,j,n}\alpha_{ij}^{k}\alpha_{nj}^{k}\sqrt{\gamma_{ij}\gamma_{nj}}\epsilon_j \ket{i}\bra{n},\\
&L_{k}^{(r)\dagger}L_{k}^{(r)}H_0 = \sum_{i,j,n}\alpha_{nj}^{k}\alpha_{ij}^{k}\sqrt{\gamma_{nj}\gamma_{ij}}\epsilon_i \ket{i}\bra{n},\\
&H_0L_{k}^{(r)\dagger}L_{k}^{(r)} = \sum_{i,j,n} \alpha_{nj}^{k}\alpha_{ij}^{k}\sqrt{\gamma_{nj}\gamma_{ij}}\epsilon_n\ket{n}\bra{i}, 
\end{split}
\end{equation}
where we can use the fact that the Lindblad operators produce jumps only between levels with a fixed energy gap $\Delta \epsilon_k = \pm \Delta \epsilon_r$, determined by the reservoir, to rewrite the $\alpha$-terms as delta functions:
\begin{equation}\label{eq:delta functions1}
\begin{split}
&\alpha_{ij}^{k}\alpha_{nj}^{k} = \alpha_{nj}^{k}\alpha_{ij}^{k} = \delta(\Delta\epsilon_{ji}-\Delta\epsilon_k)\delta(\Delta\epsilon_{jn}-\Delta\epsilon_k),
\end{split}
\end{equation}
where $\Delta \epsilon_{i j} = \epsilon_i - \epsilon_j$. 
Replacing Eq.~\eqref{eq:delta functions1} into Eq.~\eqref{eq: terms_heat} we have:
\begin{equation}\label{eq:heat_derivation_terms}
\begin{split}
    &L_{k}^{(r)\dagger}H_{0}L_{k}^{(r)} = \sum_{i,j,n} \sqrt{\gamma_{ij}\gamma_{nj}}\epsilon_{j} \ket{i}\bra{n} \delta(\Delta\epsilon_{ji}-\Delta\epsilon_{k})\delta(\Delta\epsilon_{ni}),\\
    &L_{k}^{(r)\dagger}L_{k}^{(r)}H_{0} = \sum_{i,j,n} \sqrt{\gamma_{ij}\gamma_{nj}}\epsilon_{n} \ket{i}\bra{n} \delta(\Delta\epsilon_{ji}-\Delta\epsilon_{k})\delta(\Delta\epsilon_{ni}),\\
    &H_0 L_{k}^{(r)\dagger}L_{k}^{(r)} = \sum_{i,j,n} \sqrt{\gamma_{ij}\gamma_{nj}}\epsilon_{n} \ket{n}\bra{i} \delta(\Delta\epsilon_{ji}-\Delta\epsilon_{k})\delta(\Delta\epsilon_{ni}),\\
\end{split}
\end{equation}
where we used $\delta(\Delta\epsilon_{ji}-\Delta\epsilon_{k}) \delta(\Delta\epsilon_{jn}-\Delta\epsilon_{k}) = \delta(\Delta\epsilon_{ji}-\Delta\epsilon_{k}) \delta(\Delta\epsilon_{ji}-\Delta\epsilon_{j n}) = \delta(\Delta\epsilon_{ji}-\Delta\epsilon_{k}) \delta(\Delta \epsilon_{n i})$.

By introducing the expressions in \eqref{eq:heat_derivation_terms} into \eqref{eq:heat_derivation} we arrive at a general expression for the heat currents valid for both degenerate and non-degenerate level structures in the machine: 
\begin{equation}
\begin{split}
    \langle \dot{Q}_{r} \rangle = \sum_{k,n,i,j} ^{\in B_r}\sqrt{\gamma_{ij}\gamma_{nj}}\delta(\Delta\epsilon_{ji}-\Delta\epsilon_{k})\delta(\Delta\epsilon_{ni})\\ \times [\epsilon_{j}\pi_{ni}-\epsilon_n\mathrm{Re}(\pi_{ni})]. 
\end{split}
\end{equation}

If we now particularize the above expression for the case where we don't have degenerate energy levels, the term $\delta(\epsilon_{n i})$ leads to select indices with $n= i$ and we arrive to:    
\begin{equation}\label{eq:heat_nondegenerate}
\begin{split}
\langle\dot{Q}_{r}\rangle =& \sum_{k,i,j}^{\in B_r}\delta(\Delta\epsilon_{ji}-\Delta\epsilon_{k})(\epsilon_j-\epsilon_i)\gamma_{ij}\pi_{ii} \\
&=\sum_{i<j}^{\in B_r} (\epsilon_j-\epsilon_i)(\gamma_{ij}\pi_{ii}-\gamma_{ji}\pi_{jj}),\\
\end{split}
\end{equation}
as given in Eq.~\eqref{heat_normal}. On the other hand, for the cases with degenerate energy levels, $\delta (\Delta \epsilon_{n i})$ can be zero even for $n \neq i$, and extra terms are obtained (for more details see the derivation of equations of motion in Appendix \ref{appendix: classical equivalent NIC}). In the case of a single pair of degenerate levels $\ket{\mathrm{u}}$ and $\ket{\mathrm{v}}$ we obtain:  
\begin{equation}\label{eq:heat_degenerate}
\begin{split}
\langle\dot{Q}_{r}\rangle = &\sum_{i<j}^{\in B_r} (\epsilon_j-\epsilon_i)(\gamma_{ij}\pi_{ii}-\gamma_{ji}\pi_{jj}) \\&+2 \sum_{j}^{\in B_r} (\epsilon_j-\epsilon_{\mathrm{v}})\sqrt{\gamma_{\mathrm{u}j}\gamma_{\mathrm{v}j}}\mathrm{Re}(\pi_{\mathrm{uv}}),
\end{split}
\end{equation}
as we reported in Eq.~\eqref{eq:NICheat}. Notice above that the extra term in the heat current, which is associated to transitions to or from the degenerate levels, is indeed non-zero only in the presence of coherence between the degenerate pair $\pi_{\mathrm{u v}} \neq 0$. Moreover, the coherence needs to have a real component, as it is the case of the noise-induce-coherence machine, c.f. \eqref{eq:steady_noise-induced_coherence}. On the other hand, in the case of autonomous refrigerators showing Hamiltonian-induced coherence, the coherence between degenerate levels in the steady state is a pure imaginary number [c.f.~Eq. \eqref{steady_state_coherence}], and the heat current hence reduces to the standard expression for non-degenerate levels, as in Eq.~\eqref{eq:heat_nondegenerate}. 

\section{Details on the construction of classical equivalent models}
\label{appendix: classical equivalent NIC}

In this Appendix we provide details on the procedure followed to obtain a classical thermodynamically equivalent model, in both cases of Hamiltonian-induced and noise-induced coherences. Moreover, we show that, even in the case of noise-induced coherence, it is sufficient that the classical analogue reproduces the steady-state populations in order to produce exactly the same steady state average currents. 

We start by showing how to remove the possible time dependence of the Hamiltonian in the (Hamiltonian-induced coherence) case in which we have a driving of the form \eqref{eq:driving_exp}. This is accomplished by moving to a rotating frame that simplifies the equations of motion. The rotated operators are defined by the transformation $A^{\prime} = U(t)AU^{\dagger}(t)$ for any generic operator $A$, with $U(t) = e^{\rm{i} X t}$ and $X = \sum_i (\epsilon_i +\delta_{ui} \omega_{\rm d})\ket{i}\bra{i}$. In the rotating frame, the master equation \eqref{quantum_master_equation} takes the form: 
\begin{equation}
    \dfrac{d\rho^{\prime}}{dt} = -\rm i \left [ V_I, \rho^{\prime}\right] + \sum_{r}^{R} \sum_{k = \downarrow, \uparrow} \mathcal{D}^{(r)}_{k}[\rho^\prime],
\end{equation}
with the Hamiltonian contribution $V_I = -\Delta_{\rm d} \ket{\rm u}\bra{\rm u} + g(\ket{\rm u}\bra{\rm v} + \rm{h.c})$. In the case of resonant driving or for internal Hamiltonian interactions within the machine, the rotating frame reduces to the interaction picture, with $V_I= g(\ket{\rm u}\bra{\rm v} + \rm{h.c})$.

We obtain the equations of motion for the density operator elements from the above master equation in Lindblad form, from which we can obtain the expression for a generic element of the density operator: 
\begin{equation}
\begin{split}
    \dot{\rho}_{ij} = -\mathrm{i} \bra{i}[V_I,\rho]\ket{j} + \sum_{r}\sum_{k} \bra{i}\mathcal{D}_{k}^{(r)}[\rho]\ket{j},
\end{split}
\end{equation}
where for the ease of simplicity in the notation, we ignore the prime symbols on the density matrix elements. After expanding the form of the dissipators $\mathcal{D}_k^{(r)}[\rho] = L_k^{(r)} \rho L_k^{(r) \dagger} - \frac{1}{2}\{L_k^{(r) \dagger} L_k^{(r)}, \rho\}$, we obtain the following terms contributing to the above expression:
\begin{align}\label{eq:terms}
    \bra{i}[V_I,\rho]\ket{j} =& \Delta_{\rm d} (\delta_{j\rm u}\rho_{i\rm u}+\delta_{i\rm u}\rho_{\rm u j}) \nonumber + g (\delta_{i\mathrm{u}}\rho_{\mathrm{v}j} \\ &+ \delta_{i\mathrm{v}}\rho_{\mathrm{u}j} - \delta_{\mathrm{v}j}\rho_{i\mathrm{u}}- \delta_{\mathrm{u}j}\rho_{i\mathrm{v}}), \nonumber \\
    \bra{i}L_{k}^{(r)}\rho L_{k}^{(r)\dagger}\ket{j} =& \sum_{n,m}\alpha_{ni}^{k}\alpha_{mj}^{k}\sqrt{\gamma_{ni}\gamma_{mj}}\rho_{nm},\nonumber \\
    \bra{i}L_{k}^{(r)\dagger}L_{k}^{(r)}\rho\ket{j} =& \sum_{n,m}\alpha_{im}^{k}\alpha_{nm}^{k}\sqrt{\gamma_{im}\gamma_{nm}}\rho_{nj}, \\
    \bra{i}\rho L_{k}^{(r)\dagger}L_{k}^{(r)}\ket{j} =& \sum_{n,m}\alpha_{nm}^{k}\alpha_{jm}^{k}\sqrt{\gamma_{nm}\gamma_{jm}}\rho_{in}, \nonumber 
\end{align}
where we can again use the fact that the Lindblad operators produce jumps only between levels with a fixed energy gap $\Delta \epsilon_k = \pm \Delta \epsilon_r$ and rewrite the $\alpha$-terms as delta functions:
\begin{equation}\label{eq:delta functions2}
\begin{split}
&\alpha_{ni}^{k}\alpha_{mj}^{k} = \delta(\Delta\epsilon_{in}-\Delta\epsilon_k)\delta(\Delta\epsilon_{jm}-\Delta\epsilon_k), \\&\alpha_{im}^{k}\alpha_{nm}^{k} =\delta(\Delta\epsilon_{mi}-\Delta\epsilon_{k})\delta(\Delta\epsilon_{ni}),\\&\alpha_{nm}^{k}\alpha_{jm}^{k} =\delta(\Delta\epsilon_{mn}-\Delta\epsilon_{k})\delta(\Delta\epsilon_{nj}).
\end{split}
\end{equation}

In the case of Hamiltonian-induced coherence the $\delta$ functions in \eqref{eq:delta functions2} can be simplified by taking into account the fact that in this case different transitions cannot have the same $\Delta\epsilon_k$ associated with them. 
Then \eqref{eq:delta functions2} becomes: 
\begin{equation}
\begin{split}
    &\alpha_{ni}^{k}\alpha_{mj}^{k} = \delta(\Delta\epsilon_{in}-\Delta\epsilon_k)\delta_{ij}\delta_{nm}, \\
    &\alpha_{im}^{k}\alpha_{nm}^{k} =\delta(\Delta\epsilon_{mi}-\Delta\epsilon_{k})\delta_{ni},\\
    &\alpha_{nm}^{k}\alpha_{jm}^{k} =\delta(\Delta\epsilon_{mn}-\Delta\epsilon_{k})\delta_{nj},    
\end{split}
\end{equation}
which results in the set of equations \eqref{eq._of_Motion}.  

In the case of noise-induced coherence, we don't have an interaction Hamiltonian, and therefore $V_I=0$ in the first expression on \eqref{eq:terms}, which doesn't need to be considered. The $\delta$ functions in \eqref{eq:delta functions2}, on the other hand, have more terms, since there are now different transitions associated with the same $\Delta\epsilon_k$. Now \eqref{eq:delta functions2} reads: 
\begin{align}\label{eq: simp noise-induced}
    \alpha_{ni}^{k}\alpha_{mj}^{k} =& \delta(\Delta\epsilon_{in}-\Delta\epsilon_k)(\delta_{ij}\delta_{nm} + \delta_{i\mathrm{u}}\delta_{j\mathrm{v}}\delta_{nm} + \delta_{i\mathrm{v}}\delta_{j\mathrm{u}}\delta_{nm}  \nonumber \\&+\delta_{n\mathrm{u}}\delta_{m\mathrm{v}}\delta_{ij}+\delta_{n\mathrm{v}}\delta_{m\mathrm{u}}\delta_{ij}), \\
    \alpha_{im}^{k}\alpha_{nm}^{k} =&\delta(\Delta\epsilon_{mi}-\Delta\epsilon_{k})(\delta_{ni} + \delta_{i\mathrm{v}}\delta_{n\mathrm{u}} + \delta_{i\mathrm{u}}\delta_{n\mathrm{v}}),\nonumber \\
    \alpha_{nm}^{k}\alpha_{jm}^{k} =&\delta(\Delta\epsilon_{mn}-\Delta\epsilon_{k})(\delta_{nj}+ \delta_{j\mathrm{v}}\delta_{n\mathrm{u}} + \delta_{j\mathrm{u}}\delta_{n\mathrm{v}}).\nonumber 
\end{align}
In view of \eqref{eq: simp noise-induced} the equations of motion are:
\begin{equation} \label{eqs:nicpop}
    \dot \rho_{nn} = \sum_{i} \left ( \gamma_{in}\rho_{ii} - \gamma_{ni}\rho_{nn} \right ) + 2 \sqrt{\gamma_{\text{u}n}\gamma_{\text{v}n}}\text{Re}(\rho_{\text{uv}}),
\end{equation}
for the levels $n \neq \{\text{u},\text{v}\}$, and:
\begin{equation}
\begin{split} \label{eq._of_Motion_2}
    \dot \rho_{nn} =& \sum_{i} \left [ \gamma_{in}\rho_{ii} - \left ( \gamma_{ni}\rho_{nn} + \sqrt{\gamma_{\text{u}i}\gamma_{\text{v}i}}\text{Re}(\rho_{\text{uv}}) \right) \right ],\\
    \dot \rho_{\text{uv}} =& \sum_{i} [ 2 \sqrt{\gamma_{i\text{u}}\gamma_{i\text{v}}}\rho_{ii} - \sqrt{\gamma_{\text{u}i}\gamma_{\text{v}i}} (\rho_{\text{uu}} + \rho_{\text{vv}})\\ &- (\gamma_{\text{u}i} + \gamma_{\text{v}i}) \rho_{\text{uv}}], \\
\end{split}
\end{equation}
for $n = \{\text{u},\text{v}\}$. They result in a steady-state coherence term given by \eqref{eq:steady_noise-induced_coherence}. The introduction of \eqref{eq:steady_noise-induced_coherence} into \eqref{eq._of_Motion_2} leads to corrections for the rates and a new stochastic transition between levels $\mathrm{u}$ and $\mathrm{v}$ as shown in Eq. ~\eqref{correccion_NI_CA}. 

Finally, it can be shown that all energy currents in the quantum machine and the classical equivalent are equal, also in the case of noise-induced coherence. This is a consequence of the fact that in the steady state, the value of the coherence can be written entirely in terms of the steady-state populations, as given in Eq. ~\eqref{eq:steady_noise-induced_coherence}. That leads to the substitution of the original collective transitions by independent ones with modified rates, Eq. ~\eqref{correccion_NI_CA}. As a consequence, the expression for the original heat currents in the collective transitions, Eq. ~\eqref{eq:NICheat}, can be also expressed as a function of the new transitions using the same steady-state populations, therefore guaranteeing the same average currents in the original and classical thermodynamic equivalent models.

\section{Full Counting Statistics methods}\label{appendix: full counting}

The Full Counting Statistics (FCS) formalism is used to compute the variances of the different input and output currents of the quantum thermal machines presented and their classical equivalent models. In this formalism, a set of counting fields $\vec{\chi}:=\{\chi_r\}$ is introduced that keep track of the energy quanta exchanges between the machine and the thermal reservoirs $r$. These lead to the derivation of a generalized master equation for an extended density operator $\rho_G(t,\{\chi_r\}$  ) depending on the fields. It reads~\cite{Esposito09}: 
\begin{align}\label{quantum_master_equation_FCS}
    \dfrac{d}{dt}\rho_{\mathrm{G}}(t, \{\chi_{r}\}) &= -i \left [ H(t), \rho_{\mathrm{G}}(t, \{\chi_{r}\})\right]  \nonumber \\ &+ \sum_{r=1}^R \sum_k \bar{\mathcal{D}}^{(r)}_{k}[\rho_{\mathrm{G}}(t, \{\chi_{r}\})],
\end{align}
with a new set of dissipators with a modified form: 
\begin{equation}\label{dissipator_FCS}
\begin{split}
    \bar{\mathcal{D}}^{(r)}_{k}\left[\rho_{\mathrm{G}}\right] :=& e^{-\nu_{k}^{(r)} \chi_{r}} ~ L_{k}^{(r)} \rho_{\mathrm{G}} L_{k}^{(r)~\dagger} \\ &- \dfrac{1}{2} \left \{L_{k}^{(r) \dagger}L^{(r)}_{k}, \rho_{\mathrm{G}} \right \}.  
\end{split}
\end{equation}
where the numbers $\nu_{k}^{(r)}$ are chosen to be $1$ for operators $L_k^{(r)}$ associated with the emission of a quanta into the reservoir $r$ ($\Delta \epsilon_k = - \Delta \epsilon_r$) and $-1$ for operators associated with the absorption of a quanta ($\Delta \epsilon_k = \Delta \epsilon_r$). In this way, the counting fields $\chi_r$ are associated to the net flux of quanta $\dot{N}_r$ transferred from the reservoir into the machine. 

In any case, in the limit $\{\chi_{r}\}\rightarrow 0$, we recover $\rho_{\mathrm{G}}(t) = \rho(t)$ and \eqref{quantum_master_equation_FCS} reduces to the standard master equation \eqref{quantum_master_equation} for the machine evolution. Moreover, as in the case of the original master equation, Eq.~\eqref{quantum_master_equation_FCS} can be linearized and written in the form 
\begin{equation} \label{eq:FCSeqs}
d\vec{p}_{G}(t)/dt = W_{G}(\{\chi_{r}\})\vec{p}_{G}(t), 
\end{equation}
where $\vec{p}_{G}(t)$ contains all the density operator elements and $W_G$ is a matrix capturing the dependence between elements $\rho_{ij}$ within the set of equations of motion. We are interested in the eigenvalue $\lambda(\{ \chi_r\})$ of the matrix $W_G$ with the largest real part, which is related to the machine cumulant generating function $\mathcal{K}({\chi_r},t)$. Indeed for systems with a single steady state we have that in the long time limit~\cite{Schaller14}:
\begin{equation}
    \mathcal{K}(\{\chi_r\}, t) \rightarrow \lambda(\{\chi_r\})t.
\end{equation}
Then the cumulants $\mathcal{C}_n^{(r)}$ associated to the exchange of quanta with the different reservoirs corresponding to the counting fields $\chi_r$, can be obtained as derivatives with respect to that counting fields of this eigenvalue, evaluated for all fields equal to zero: 
\begin{equation}
    \mathcal{C}_n^{(r)} = (-\mathrm{i}\partial_{\chi_r})^{n} \lambda(\{\chi_{l}\}) |_{\{\chi_{l}\} = 0}, 
\end{equation}
for $n=1, 2, 3, ...$. Here the first ($n=1$) and second ($n=2$) cumulants correspond, respectively, to the average ($\mathcal{C}_1^{(r)} = \langle \dot{N}_r \rangle$) and variance ($\mathcal{C}_2^{(r)} = \mathrm{Var}[\dot{N}_r]$) of the currents of quanta on that reservoir. The average and variances of the heat currents in which we are mainly interested in this work are then given by: 
\begin{equation}
    \langle \dot{Q}_r \rangle = \Delta \epsilon_r \mathcal{C}_1^{(r)}, ~~~~~~ \mathrm{Var}[\dot{Q}_r] = \Delta \epsilon_r^2 \mathcal{C}_2^{(r)}.
\end{equation}
Due to the preservation of quanta exchanges in all the interactions in the weak coupling limit, it follows that for long trajectories in steady-state machines where the different reservoirs contribute the same number of quanta along all possible cycles, we have a proportionality among all currents. That means that the fluctuations of every current are also proportional:
\begin{equation} \label{eq:unique_current}
\begin{split}
\mathrm{Var}[\dot{Q}_r] &= \Delta\epsilon_r^{2}\mathrm{Var} [\dot{N}], \\
\mathrm{Cov}[\dot{Q}_r \dot{Q}_s] &= \Delta\epsilon_r \Delta\epsilon_s \mathrm{Var} [\dot{N}],    
\end{split}
\end{equation}
where ${\rm Var}[\dot{N}] = {\rm Var}[\dot{N}_r]$ for all $r$. Analogously, the average and variance of the power for the case of external driving are given, respectively, by the first law, $\langle \dot{W} \rangle = \sum_{r} \langle \dot{Q}_r \rangle$, and in the long-time limit we have: 
\begin{equation} \label{eq:work_variance}
\begin{split}
\mathrm{Var}[\dot{W}] &= \sum_{r} \left( \mathrm{Var}[\dot{Q}_r] + 2 \sum_s \mathrm{Cov}[\dot{Q}_r \dot{Q}_s] \right) \\ 
&= \omega_d^{2} \mathrm{Var} [\dot{N}].    
\end{split}
\end{equation}
where we have used Eqs.~\eqref{eq:unique_current}.

Unfortunately the size and complexity of the matrix $W_G$ in many cases makes impossible to obtain analytically the largest eigenvalue $\lambda(\{\chi_r\})$ by direct diagonalization of $W_G$, and other methods are required. In order to obtain the first and second cumulants analytically, we follow the method known as ``Inverse Counting Statistics'' originally introduced in Ref.~\cite{Bruderer14} and used recently in Refs.~\cite{Kalaee21,Manzano23} for similar purposes. In the following we review this method for the case of a single field $\chi$, which is enough for obtaining all the relevant quantities in our case, but the expressions can be naturally extended to multiple fields $\{ \chi_r\}$ (see e.g. Appendix C in Ref.~\cite{Manzano23}).

In this method, the characteristic polynomial of $W_{G}$, namely, $\mathrm{Pol(\lambda)}:= -\mathrm{det}[W_{G}(\chi)-\lambda \mathds{1}]$, is expanded in series in terms of the powers of its eigenvalues: 
\begin{equation}
   \mathrm{Pol}(\lambda) = \sum_{n = 0}^{M} a_{n}(\chi) \lambda^{n}(\chi) = 0, 
\end{equation}
where $M$ is the range of the matrix $W_G$. Now we define the coefficients:
\begin{align}
    a_{n}^{\prime} &=\mathrm{i} \partial_{\chi}a_{n}|_{\chi=0}, \\
    a_{n}^{\prime\prime} &=(\mathrm{i} \partial_{\chi})^{2}a_{n}|_{\chi=0} = -\partial_{\chi}^{2}a_{n}|_{\chi=0}, 
\end{align}
and similarly denote $\lambda^\prime = i \partial_\chi \lambda|_{\chi = 0}$ and $\lambda^{\prime \prime} = -\partial^2_\chi|_{\chi = 0}$. The first derivative of the entire characteristic polynomial is then given by:
\begin{equation}\label{eq: first derivative}
    \left[ \mathrm{i} \partial_{\chi}\sum_{n}^{M} a_{n}\lambda^{n}\right]_{\chi=0} = \sum_{n}^{M} \left[a_{n}^{\prime} + (n+1)a_{n+1}\lambda^{\prime}\right]\lambda^{n}(0), 
\end{equation}
and the second derivative reads:
\begin{equation}\label{eq_ second derivative}
\begin{split}
    &\left[(-\mathrm{i}\partial_{\chi})^{2}\sum_{n=0}^{M}a_{n}\lambda^{n}\right]_{\chi=0} = \sum_{n=0}^{M}[ a^{\prime\prime}_{n} + 2(n+1) a^{\prime}_{n+1}\lambda^{\prime} \\
    & + (n+1)a_{n+1}\lambda^{\prime\prime} + (n+1)(n+2)a_{n+2}\lambda^{\prime 2}]\lambda^{n}(0).
\end{split}
\end{equation}
Since Pol($\lambda$)$= 0$, both equations above should be equal to zero. Therefore, if the system has a unique steady state, such that $\lambda(0) = 0$, as it is our case, then the zero order term in $\lambda$ vanish, and we obtain from \eqref{eq: first derivative}:  
\begin{equation}
    a_{0}^{\prime} + a_{1}\lambda^{\prime} = 0, 
\end{equation}
so that the first cumulant (average current) is given by: 
\begin{equation} \label{eq:cumulant1}
    \mathcal{C}_1 = \lambda^{\prime} = - \dfrac{a_{0}^{\prime}}{a_{1}},
\end{equation}
and in the same way from \eqref{eq_ second derivative} we obtain the second cumulant (variance): 
\begin{equation} \label{eq:cumulant2}
    \mathcal{C}_2 = -\dfrac{1}{a_1}\left ( a_{0}^{\prime\prime}+ 2a_1^{\prime}\mathcal{C}_1 + 2a_2 \mathcal{C}_1^2 \right).
\end{equation}
Replacing the expression for the cumulant $\mathcal{C}_1$ in Eqs.~\eqref{eq:cumulant1} into the above Eq.~\eqref{eq:cumulant2}, we obtain the final expression for the variance of the probability current in the long-time limit:
 \begin{equation}
  {\rm Var}[\dot{N}] = -\frac{1}{a_1}[a_0^{''} - 2a_1^{'} \Big(\frac{a_0^{'}}{a_1}\Big) + 2a_2 \Big(\frac{a_0^{'}}{a_1}\Big)^2],  
\end{equation}
which is used to obtain all the variances for heat and work currents [Eqs.~\eqref{eq:unique_current} and \eqref{eq:work_variance}].\\

\section{Expressions from Inverse Counting Statistics}\label{appendix: cumulants}

In this Appendix we provide the explicit form of the FCS matrix $W_{G}$ and the relevant coefficients of the characteristic polynomial used to obtain analytical expressions for the currents variances using the Inverse Counting Statistics method, as introduced in Appendix \ref{appendix: full counting}. For convenience we rewrite here the expressions of the probability flux in the long-time limit in terms of the relevant coefficients for both the quantum machines and their classical-equivalent counterparts:
 \begin{equation}\label{eq:var_with_coef}
 \begin{split}
  {\rm Var}[\dot{N}] = -\frac{1}{a_1}[a_0^{''} - a_1^{'} \Big(\frac{a_0^{'}}{a_1}\Big) + a_2 \Big(\frac{a_0^{'}}{a_1}\Big)^2],  \\
  {\rm Var}[\dot{N}^{\rm cl}] = -\frac{1}{a_1^{\rm cl}}[a_0^{{\rm cl}''} - a_1^{{\rm cl}'} \Big(\frac{a_0^{{\rm cl}'}}{a_1^{\rm cl}}\Big) + a_2^{\mathrm{cl}} \Big(\frac{a_0^{{\rm cl}'}}{a_1^{\rm cl}}\Big)^2],
 \end{split} 
\end{equation}
which are used to evaluate the variance of the different heat currents and power, as well as the fluctuations ratio $\mathcal{R}$ as introduced in Eq.~\eqref{eq:R}. 

In the following subsections we first provide the expressions for the general cases of $N$-level multi-cycle machines discussed in the Results section for Hamiltonian-induced coherence and noise-induced coherence, together with their corresponding classical equivalents. Then we also include the particular expressions for the three quantum thermal machines used as main illustrative examples. We stress that while the explicit final expressions of ${\rm Var}[\dot{N}]$ and ${\rm Var}[\dot{N}^{\rm cl}]$ as well as those of some of the coefficients are not included here for size reasons, all analytical expressions can be found in the supplemental repository \cite{SM_repository}.

\subsection{Generic multi-cycle Hamiltonian-induced coherence machines}

Following the coarse-graining procedure described in the main text we obtain the FCS matrix $W_G(\chi)$ in Eq. ~\eqref{eq:FCSeqs} with a single counting field $\chi$ for quantum machines with Hamiltonian-induced coherence in one of their transitions. Recall that this matrix includes only the relation between the relevant coherences and the diagonal elements of the density matrix in the steady state, while those associated to coherences that become zero in the steady-state are not included. In the case of Hamiltonian-induced coherence this is the case of the coherent subspace (spanned by states $\ket{u}$ and $\ket{v}$), while coherences related to the monitoring state $\ket{m}$ or the coarse-grained state $\ket{S}$ (involving the rest of the system populations) are not included. Taking that into account, the matrix $W_G$ associated to the vector $\vec{p}_{G}(t) = (\rho_{\rm uu},\rho_{\rm vv},\rho_{\rm mm},\rho_{\rm ss}, \mathrm{Re}[\rho_{\rm{uv}}],\mathrm{Im}[\rho_{\rm uv}])$ has the form 
\begin{equation}
W_{G}(\chi) =\left(
\begin{array}{cccccc}
 -\gamma _{\text{vm}}-\Gamma _{\text{vs}} & 0 & \gamma _{\text{mv}} & \Gamma _{\text{sv}} & 0 & -2 g \\
 0 & -\gamma _{\text{um}}-\Gamma _{\text{us}} & e^{-i \chi } \gamma _{\text{mu}} & \Gamma _{\text{su}} & 0 & 2 g \\
 \gamma _{\text{vm}} & e^{i \chi } \gamma _{\text{um}} & -\Gamma _{\text{ms}}-\gamma _{\text{mu}}-\gamma _{\text{mv}} & \Gamma _{\text{sm}} & 0 & 0 \\
 \Gamma _{\text{vs}} & \Gamma _{\text{us}} & \Gamma _{\text{ms}} & -\Gamma _{\text{sm}}-\Gamma _{\text{su}}-\Gamma _{\text{sv}} & 0 & 0 \\
 0 & 0 & 0 & 0 & -\Lambda  & \Delta _d \\
 g & -g & 0 & 0 & -\Delta _d & -\Lambda  \\
\end{array}
\right), 
\end{equation}
where in the last row standing for the imaginary part of the coherence between states $\ket{\rm u}$ and $\ket{\rm v}$, we defined $\Lambda = (\Gamma_{\mathrm{vs}}+\Gamma_{\mathrm{us}}+\gamma_{\rm{vm}}+\gamma_{\mathrm{um}})/2$. The most relevant inverse counting coefficients associated with this matrix are (the rest can be found in the supplemental repository\cite{SM_repository}):  
{\small
\begin{align} \label{eq:terms_hamiltonian_induced}
        a_{0}^{\prime} = &\Gamma_{\text{ms}} \gamma_{\text{um}} \left(\Gamma_{\text{su}} \left[\left(\Delta _d^2+\Lambda ^2\right) \left(\gamma_{\text{vm}}+\Gamma_{\text{vs}}\right)+2 g^2 \Lambda \right] + 2 g^2 \Lambda  \Gamma_{\text{sv}}\right) +  \gamma_{\text{mv}} \gamma_{\text{um}} (\Gamma_{\text{su}} \left[\Gamma_{\text{vs}} \left(\Delta _d^2+\Lambda ^2\right)
        +2 g^2 \Lambda \right] \nonumber \\
        &+2 g^2 \Lambda  \Gamma_{\text{sm}} +2 g^2 \Lambda  \Gamma_{\text{sv}})
        -\gamma_{\text{mu}} [\gamma_{\text{vm}} \left[\Delta_d^2 \Gamma_{\text{us}} \left(\Gamma_{\text{sm}}+\Gamma_{\text{sv}}\right)+\Lambda  \left(2 g^2 \left(\Gamma_{\text{sm}}+\Gamma_{\text{su}}+\Gamma_{\text{sv}}\right)+\Lambda  \Gamma_{\text{us}} \left(\Gamma_{\text{sm}}+\Gamma_{\text{sv}}\right)\right)\right] \nonumber \\ 
        &+\Gamma_{\text{sm}} \left(\Gamma_{\text{us}} \left(\Gamma_{\text{vs}} \left(\Delta_d^2+\Lambda ^2\right)+2 g^2 \Lambda \right)+2 g^2 \Lambda  \Gamma_{\text{vs}} \right)], \\
        a_{0}^{\prime\prime} = & -\Gamma_{\text{ms}} \gamma_{\text{um}} \left(\Gamma_{\text{su}} \left[\left(\Delta_d^2+\Lambda ^2\right) \left(\gamma_{\text{vm}}+\Gamma_{\text{vs}}\right)+2 g^2 \Lambda \right]+2 g^2 \Lambda  \Gamma_{\text{sv}}\right) -\gamma_{\text{mv}} \gamma_{\text{um}} (\Gamma_{\text{su}} \left[\Gamma_{\text{vs}} \left(\Delta_d^2+\Lambda ^2\right)+2 g^2 \Lambda \right] \nonumber\\
        &+2 g^2 \Lambda  \Gamma_{\text{sm}}+2 g^2 \Lambda  \Gamma_{\text{sv}}) 
        -\gamma_{\text{mu}} [\gamma_{\text{vm}} \left(\Delta_d^2 \Gamma_{\text{us}} \left(\Gamma_{\text{sm}}+\Gamma_{\text{sv}}\right)+\Lambda  \left[2 g^2 \left(\Gamma_{\text{sm}}+\Gamma_{\text{su}}+\Gamma_{\text{sv}}\right)+\Lambda  \Gamma_{\text{us}} \left(\Gamma_{\text{sm}}+\Gamma_{\text{sv}}\right)\right]\right) \nonumber \\
        & +\Gamma_{\text{sm}} \left(\Gamma_{\text{us}} \left(\Gamma_{\text{vs}} \left(\Delta_d^2+\Lambda ^2\right)+2 g^2 \Lambda \right)+2 g^2 \Lambda  \Gamma_{\text{vs}}\right)], \nonumber \\
        a_{1}^{\prime} = & 2 \gamma_{\text{mv}} \gamma_{\text{um}} \left(g^2 \left(\Lambda +\Gamma_{\text{sm}}+\Gamma_{\text{su}}+\Gamma_{\text{sv}}\right)+ \Lambda  \Gamma_{\text{su}} \Gamma_{\text{vs}}\right) + \Gamma_{\text{ms}} \gamma_{\text{um}} \left(\Gamma_{\text{su}} \left[\Delta_d^2+2 g^2+\Lambda^2+2 \Lambda  \left(\gamma_{\text{vm}}+\Gamma_{\text{vs}}\right)\right]+ 2 g^2 \Gamma_{\text{sv}}\right) \nonumber \\
        &-\gamma_{\text{mu}} \left(\Gamma_{\text{sm}} \left(\Gamma_{\text{us}} \left(\Delta_d^2+2 g^2+\Lambda^2+2 \Lambda  \Gamma_{\text{vs}}\right)+2 g^2 \Gamma_{\text{vs}}\right)+2 \gamma_{\text{vm}} \left[g^2 \left(\Lambda +\Gamma_{\text{sm}}+\Gamma_{\text{su}}+\Gamma_{\text{sv}}\right)+\Lambda  \Gamma_{\text{us}} \left(\Gamma_{\text{sm}}+\Gamma_{\text{sv}}\right)\right]\right). \nonumber
\end{align}}

For the classical equivalent of this model we no longer have the coherent contribution, but a new stochastic transition between the two levels involved in the coherent interaction with rates $\gamma_{\rm{vu}}^{\rm cl} = \gamma_{\rm{uv}}^{\rm cl}$ as given in Eq. ~\eqref{classical_rate}. The matrix $W_{G}^{\mathrm{cl}}$ in this case reads: 
\begin{equation}
W_{G}^{\mathrm{cl}}(\chi) = \left(
\begin{array}{cccc}
 -\gamma_{\text{vu}}{}^{\text{cl}}-\gamma_{\text{vm}}-\Gamma_{\text{vs}} & \gamma_{\text{vu}}{}^{\text{cl}} & \gamma_{\text{mv}} & \Gamma_{\text{sv}} \\
 \gamma_{\text{vu}}{}^{\text{cl}} & -\gamma_{\text{vu}}{}^{\text{cl}}-\gamma_{\text{um}}-\Gamma_{\text{us}} & e^{-i \chi } \gamma_{\text{mu}} & \Gamma_{\text{su}} \\
 \gamma_{\text{vm}} & e^{i \chi } \gamma_{\text{um}} & -\Gamma_{\text{ms}}-\gamma_{\text{mu}}-\gamma_{\text{mv}} & \Gamma_{\text{sm}} \\
 \Gamma_{\text{vs}} & \Gamma_{\text{us}} & \Gamma_{\text{ms}} & -\Gamma_{\text{sm}}-\Gamma_{\text{su}}-\Gamma_{\text{sv}} \\
\end{array}
\right),
\end{equation}
which is associated to the reduced vector $\vec{p}_{G}^{\rm cl}(t) = (\rho_{\rm uu},\rho_{\rm vv},\rho_{\rm mm},\rho_{\rm ss})$. The associated relevant inverse counting coefficients (all coefficients can be found in \cite{SM_repository}) are: 
\begin{equation}\label{eq:terms_hamiltonian_induced_classical}
    \begin{split}
        &a_{0}^{\mathrm{cl}\,\prime} = \dfrac{a_{0}^{\prime}}{\Lambda^2 + \Delta_{\rm d}^2}, ~~~~ a_{0}^{\mathrm{cl}\,\prime\prime} =\dfrac{a_{0}^{\prime\prime}}{\Lambda^2 + \Delta_{\rm d}^2},~~~~ a_{1}^{\mathrm{cl}\,\prime} = \frac{2 g^2 \Lambda  \left(\gamma_{\text{mv}} \gamma_{\text{um}}-\gamma_{\text{mu}} \gamma_{\text{vm}}\right)}{\Delta_d^2+\Lambda ^2}+\Gamma_{\text{ms}} \Gamma_{\text{su}} \gamma_{\text{um}}-\gamma_{\text{mu}} \Gamma_{\text{sm}} \Gamma_{\text{us}}.\\ 
    \end{split}
\end{equation}

\subsection{Three-level coherent amplifier}

To obtain the form of $W_G$ for this case, we write the system of equations for all elements of $\rho_G$ given by \eqref{quantum_master_equation_FCS}, using the Hamiltonian and Lindblad operators given in Eq. \eqref{eq: Hamiltonian Amplifier} and \eqref{eq: lindblad operator Amplifier}. We consider only the relevant elements of the matrix leading to non-zero values of the density operator $\pi$ in the steady state. That is, we include terms connecting the level populations and the imaginary part of the coherence between states $\ket{0}$ and $\ket{1}$ (see App. \ref{appendix: classical equivalent NIC}). On the other hand, both the real part of $\pi_{12}$ and the real and imaginary part of the other coherences $\pi_{13}$ and $\pi_{23}$ become zero at the steady state and we don't need to describe their evolution. The matrix reads: 
\begin{equation*}
W_{G} = \left( \begin{array}{cccc}
-\gamma_{02}  & 0 & \gamma_{20} & 2g\\
0 & -\gamma_{12} & \gamma_{21}~\mathrm{e}^{-i\chi} & -2g \\
\gamma_{02} & \gamma_{12}~\mathrm{e}^{i\chi} & -(\gamma_{20} + \gamma_{21}) & 0 \\
g & -g & 0 & -\dfrac{1}{2} (\gamma_{02}+\gamma_{12})\end{array} \right),
\end{equation*}
associated to vector $\vec{p}_{G}(t) = (\rho_{00},\rho_{11},\rho_{22},\mathrm{Im}[\rho_{01}])$. The inverse counting coefficients are: 
\begin{equation}
    \begin{split}
        &a_{0}^{\prime} = 2 g^2 \gamma _c \gamma _h \left(\bar{n}_c-\bar{n}_h\right),~~~~ a_{0}^{\prime\prime} = -2 g^2 \gamma _c \gamma _h \left(\bar{n}_c \left(2 \bar{n}_h+1\right)+\bar{n}_h\right),\\
        &a_{1} = \frac{1}{2} \left(\gamma _c \left(\bar{n}_c \left(\gamma _h^2 \bar{n}_h \left(3 \bar{n}_h+1\right)+12 g^2\right)+\gamma _h^2 \bar{n}_h^2+8 g^2\right)+\gamma _c^2 \gamma _h \bar{n}_c \left(\bar{n}_c \left(3 \bar{n}_h+1\right)+\bar{n}_h\right)+4 g^2 \gamma _h \left(3 \bar{n}_h+2\right)\right),\\
        &a_2 = \frac{1}{2} \left(\gamma _c \gamma _h \left(\bar{n}_c \left(10 \bar{n}_h+3\right)+3 \bar{n}_h\right)+\gamma _c^2 \bar{n}_c \left(2 \bar{n}_c+1\right)+\gamma _h^2 \bar{n}_h \left(2 \bar{n}_h+1\right)+8 g^2\right).
    \end{split}
\end{equation}
and $a_{1}^{\prime} = 0$. In the case of the classical equivalent, we no longer consider the contribution of coherence to the dynamics, but introduce the classical stochastic transition rate between the interacting levels, leading to a matrix $W_G^{\mathrm{cl}}$ that reads:
\begin{equation*}
W_{G}^{\mathrm{cl}} = \left( \begin{array}{ccc}
-(\gamma_{02}-\gamma_{10}^{\mathrm{cl}})  & \gamma_{10}^{\mathrm{cl}} & \gamma_{20}  \\
\gamma_{10}^{\mathrm{cl}} & -(\gamma_{12}+\gamma_{10}^{\mathrm{cl}}) & \gamma_{21} ~\mathrm{e}^{-i\chi} \\
\gamma_{02} & \gamma_{12}~\mathrm{e}^{i\chi} & -(\gamma_{20} + \gamma_{21})\end{array} \right),\\
\end{equation*}
with the corresponding associated vector $\vec{p}_{G}^{\,\mathrm{cl}}(t) = (\rho_{00},\rho_{11},\rho_{22})$. The associated inverse counting coefficients (all coefficients can be found in \cite{SM_repository}) are: 
\begin{align}
        &a_{0}^{\mathrm{cl}\,\prime} = \frac{4 g^2 \gamma _c \gamma _h \left(\bar{n}_h-\bar{n}_c\right)}{\gamma _c \bar{n}_c+\gamma _h \bar{n}_h},~~~~ a_{0}^{\mathrm{cl}\,\prime\prime} = \frac{4 g^2 \gamma _c \gamma _h \left(\bar{n}_h-\bar{n}_c\right)}{\gamma _c \bar{n}_c+\gamma _h \bar{n}_h},~~~~ a_{1}^{\mathrm{cl}\,\prime} = 0,\\
        &a_{1}^{\mathrm{cl}} = -\frac{8 g^2 \left(\gamma _c+\gamma _h\right)}{\gamma _c \bar{n}_c+\gamma _h \bar{n}_h}-\gamma _c \gamma _h \bar{n}_c-\gamma _c \gamma _h \left(3 \bar{n}_c+1\right) \bar{n}_h-12 g^2, ~~~~ a_2^{\mathrm{cl}} = -\frac{8 g^2}{\gamma _c \bar{n}_c+\gamma _h \bar{n}_h}-\gamma _c \left(2 \bar{n}_c+1\right)-\gamma _h \left(2 \bar{n}_h+1\right). \nonumber
\end{align}

\subsection{Three-qubit autonomous refrigerator}

Using the Hamiltonian and the Lindblad operators given in Eq. \eqref{eq: Hamiltonian QAR} and \eqref{eq: lindblad operators QAR}, we can derive in an analogous way the $W_G$ matrices for the autonomous absorption refrigerator and its classical equivalent. Here again, we consider only relevant elements of the density matrix, which now consist of the populations of the $8$ energy levels (in the three-qubit composed ladder) and a pure imaginary coherence between the states $\ket{101}$ and $\ket{010}$. The $W_G$ matrix now have the form:

\begin{equation*}
\begin{split}
&W_{G} =
\left(
\begin{array}{cccccccccc}
 \Gamma_1 & \gamma _{10}^{\mathrm{h}}  & \gamma _{10}^{\mathrm{c}}  \mathrm{e}^{-i \chi } & 0 & 0 & 0 & 0 & \gamma _{10}^{\mathrm{m}}   & 0 \\
\gamma _{01}^{\mathrm{h}} & \Gamma_2 & 0 & \gamma _{10}^{\mathrm{m}}  & 0 & 0 &  \gamma _{10}^{\mathrm{c}}  \mathrm{e}^{-i \chi }& 0 & 0 \\
 \gamma _{01}^{\mathrm{c}} \mathrm{e}^{i \chi } & 0 & \Gamma_3 & 0 & \gamma _{10}^{\mathrm{m}}  & 0 & \gamma _{10}^{\mathrm{h}}  & 0 & 0 \\
 0 &  \gamma _{01}^{\mathrm{m}}  & 0 & \Gamma_4 & 0 & \gamma _{10}^{\mathrm{c}}  \mathrm{e}^{-i \chi} & 0 & \gamma _{01}^{\mathrm{h}}   & 0 \\
 0 & 0 & \gamma _{01}^{\mathrm{m}}   & 0 & \Gamma_5 & \gamma _{10}^{\mathrm{h}} & 0 & \gamma _{01}^{\mathrm{c}} \mathrm{e}^{i \chi } & 0 \\
 0 & 0 & 0 & \gamma _{01}^{\mathrm{c}}  \mathrm{e}^{i \chi } & \gamma _{01}^{\mathrm{h}}   & \Gamma_6 & \gamma _{01}^{\mathrm{m}}   & 0 & 0 \\
 0 & \gamma _{01}^{\mathrm{c}}  \mathrm{e}^{i \chi } &  \gamma _{01}^{\mathrm{h}}  & 0 & 0 & \gamma _{10}^{\mathrm{m}}  & \Gamma_7 & 0 & g \\
 \gamma _{01}^{\mathrm{m}}   & 0 & 0 & \gamma _{10}^{\mathrm{h}}  & \gamma _{10}^{\mathrm{c}}  \mathrm{e}^{-i \chi } & 0 & 0 & \Gamma_8 & - g \\
 0 & 0 & 0 & 0 & 0 & 0 &  -g & g & \Gamma_9 \\
\end{array}
\right),
\end{split}
\end{equation*} 
associated to vector $\vec{p}_{G}(t) = (\rho_{000},\rho_{001},\rho_{100},\rho_{011}, \rho_{111}, \rho_{101}, \rho_{010}, \mathrm{Im}[\rho_{010-101}])$. In the above equation, we defined the aggregated rates $\Gamma_{1} = -(\gamma_{01}^{\mathrm{c}}+\gamma _{01}^{\mathrm{m}}+\gamma _{01}^{\mathrm{h}})$, $\Gamma_2 = -(\gamma_{01}^{\mathrm{c}}+\gamma _{01}^{\mathrm{m}}+\gamma _{10}^{\mathrm{h}})$, $\Gamma_3 = -(\gamma _{10}^{\mathrm{c}}+\gamma _{01}^{\mathrm{m}}+\gamma _{01}^{\mathrm{h}})$, $\Gamma_4 =  -(\gamma _{01}^{\mathrm{c}}+\gamma _{10}^{\mathrm{m}}+\gamma _{10}^{\mathrm{h}})$, $\Gamma_5 = -(\gamma _{10}^{\mathrm{c}}+\gamma _{10}^{\mathrm{m}}+\gamma _{01}^{\mathrm{h}})$, $\Gamma_{6} = -(\gamma _{10}^{\mathrm{c}}+\gamma _{10}^{\mathrm{m}}+\gamma _{10}^{\mathrm{h}})$, $\Gamma_7 = -(\gamma _{10}^{\mathrm{c}}+\gamma _{01}^{\mathrm{m}}+\gamma _{10}^{\mathrm{h}})$, $\Gamma_8 = -(\gamma _{01}^{\mathrm{c}}+\gamma _{10}^{\mathrm{m}}+\gamma _{01}^{\mathrm{h}})$ and $\Gamma_9 = -\frac{1}{2}(\gamma _d+\gamma _{2 d}+\gamma _{3 d}+\gamma _u+\gamma _{2 u}+\gamma _{3 u})$.

Some relevant inverse counting coefficients are in this case (all coefficients can be found in \cite{SM_repository}): 
{\small
\begin{align}
        a_{0}^{\prime} =& 2 g^2 \gamma _c \gamma _h \gamma _m \left(\bar{n}_h \left(\bar{n}_c+\bar{n}_m+1\right)-\bar{n}_c \bar{n}_m\right) \left(\gamma _c \left(2 \bar{n}_c+1\right)+\gamma _h \left(2 \bar{n}_h+1\right)\right)\left(\gamma _c \left(2 \bar{n}_c+1\right)+\gamma _m \left(2 \bar{n}_m+1\right)\right)\nonumber\\ & \left(\gamma _h \left(2 \bar{n}_h+1\right)+\gamma _m \left(2 \bar{n}_m+1\right)\right) \left(\gamma _c \left(2 \bar{n}_c+1\right)+\gamma _h \left(2 \bar{n}_h+1\right)+\gamma _m \left(2 \bar{n}_m+1\right)\right),\\
        a_{0}^{\prime\prime} =& 2 g^2 \gamma _c \gamma _h \gamma _m \left(\bar{n}_h \left(\bar{n}_c \left(2 \bar{n}_m+1\right)+\bar{n}_m+1\right)+\bar{n}_c \bar{n}_m\right) \left(\gamma _c \left(2 \bar{n}_c+1\right)+\gamma _h \left(2 \bar{n}_h+1\right)\right) \left(\gamma _c \left(2 \bar{n}_c+1\right)+\gamma _m \left(2 \bar{n}_m+1\right)\right)\nonumber\\& \left(\gamma _h \left(2 \bar{n}_h+1\right)+\gamma _m \left(2 \bar{n}_m+1\right)\right) \left(\gamma _c \left(2 \bar{n}_c+1\right)+\gamma _h \left(2 \bar{n}_h+1\right)+\gamma _m \left(2 \bar{n}_m+1\right)\right),\nonumber\\
        a_{1}^{\prime} =& 4 g^2 \gamma _c \gamma _h \gamma _m \left(\bar{n}_h \left(\bar{n}_c+\bar{n}_m+1\right)-\bar{n}_c \bar{n}_m\right) (4 \gamma _c^2 \left(2 \bar{n}_c+1\right){}^2 \left(\gamma _h \left(2 \bar{n}_h+1\right)+\gamma _m \left(2 \bar{n}_m+1\right)\right)+\gamma _c \left(2 \bar{n}_c+1\right) (9 \gamma _h \gamma _m \left(2 \bar{n}_h+1\right)\nonumber\\& \left(2 \bar{n}_m+1\right)+4 \gamma _h^2 \left(2 \bar{n}_h+1\right){}^2+4 \gamma _m^2 \left(2 \bar{n}_m+1\right){}^2)+\gamma _c^3 \left(2 \bar{n}_c+1\right){}^3+\left(\gamma _h \left(2 \bar{n}_h+1\right)+\gamma _m \left(2 \bar{n}_m+1\right)\right) (3 \gamma _h \gamma _m \left(2 \bar{n}_h+1\right) \nonumber\\&\left(2 \bar{n}_m+1\right)+\gamma _h^2 \left(2 \bar{n}_h+1\right){}^2+\gamma _m^2 \left(2 \bar{n}_m+1\right){}^2)).\nonumber
\end{align}}
For the classical equivalent we instead loose the coherence contribution, while adding the extra rate $\gamma^\mathrm{cl}$ between degenerated levels:
\begin{equation*}
\begin{split}
&W_{G}^{\mathrm{cl}} =\left(
\begin{array}{ccccccccc}
 \Gamma_1 & \gamma _{10}^{\mathrm{h}}  & \gamma _{10}^{\mathrm{c}}  \mathrm{e}^{-i \chi } & 0 & 0 & 0 & 0 & \gamma _{10}^{\mathrm{m}}   \\
\gamma _{01}^{\mathrm{h}}  & \Gamma_2 & 0 & \gamma _{10}^{\mathrm{m}}  & 0 & 0 &  \gamma _{10}^{\mathrm{c}}  \mathrm{e}^{-i \chi}& 0 \\
 \gamma _{01}^{\mathrm{c}} \mathrm{e}^{i \chi } & 0 & \Gamma_3 & 0 & \gamma _{10}^{\mathrm{m}} & 0 & \gamma _{10}^{\mathrm{h}}  & 0 \\
 0 &  \gamma _{01}^{\mathrm{m}}  & 0 & \Gamma_4 & 0 & \gamma _{10}^{\mathrm{c}}  \mathrm{e}^{-i \chi } & 0 & \gamma _{01}^{\mathrm{h}}   \\
 0 & 0 & \gamma _{01}^{\mathrm{m}}   & 0 & \Gamma_5 & \gamma _{10}^{\mathrm{h}}  & 0 & \gamma _{01}^{\mathrm{c}} \mathrm{e}^{i \chi } \\
 0 & 0 & 0 & \gamma _{01}^{\mathrm{c}}  \mathrm{e}^{i \chi } & \gamma _{01}^{\mathrm{h}}  & \Gamma_6 & \gamma _{01}^{\mathrm{m}}   & 0 \\
 0 & \gamma _{01}^{\mathrm{c}}  \mathrm{e}^{i \chi } &  \gamma _{01}^{\mathrm{h}} & 0 & 0 & \gamma _{10}^{\mathrm{m}}  & \Gamma_7 - \gamma^{\mathrm{cl}} & \gamma^{\mathrm{cl}}\\
 \gamma _{01}^{\mathrm{m}}   & 0 & 0 & \gamma _{10}^{\mathrm{h}}  & \gamma _{10}^{\mathrm{c}}  \mathrm{e}^{-i \chi } & 0 & \gamma^{\mathrm{cl}} & \Gamma_8 - \gamma^{\mathrm{cl}} \\
\end{array}
\right),
\end{split}
\end{equation*}
for vector $\vec{p}_{G}^{\,\mathrm{cl}}(t) = (\rho_{000},\rho_{001},\rho_{100},\rho_{011}, \rho_{111}, \rho_{101}, \rho_{010})$. 
The corresponding coefficients in the classical equivalent machine read:
{\small
\begin{align}
        a_{0}^{\mathrm{cl}\,\prime} =& -4 g^2 \gamma _c \gamma _h \gamma _m \left(\bar{n}_h \left(\bar{n}_c+\bar{n}_m+1\right)-\bar{n}_c \bar{n}_m\right) \left(\gamma _c \left(2 \bar{n}_c+1\right)+\gamma _h \left(2 \bar{n}_h+1\right)\right) \left(\gamma _c \left(2 \bar{n}_c+1\right)+\gamma _m \left(2 \bar{n}_m+1\right)\right)\nonumber \\ &~~\left(\gamma _h \left(2 \bar{n}_h+1\right)+\gamma _m \left(2 \bar{n}_m+1\right)\right),\nonumber\\
        a_{0}^{\mathrm{cl}\,\prime\prime} =& -4 g^2 \gamma _c \gamma _h \gamma _m \left(\bar{n}_h \left(\bar{n}_c \left(2 \bar{n}_m+1\right)+\bar{n}_m+1\right)+\bar{n}_c \bar{n}_m\right) \left(\gamma _c \left(2 \bar{n}_c+1\right)+\gamma _h \left(2 \bar{n}_h+1\right)\right) \left(\gamma _c \left(2 \bar{n}_c+1\right)+\gamma _m \left(2 \bar{n}_m+1\right)\right)\nonumber\\ &~ \left(\gamma _h \left(2 \bar{n}_h+1\right)+\gamma _m \left(2 \bar{n}_m+1\right)\right),\nonumber\\
        a_{1}^{\mathrm{cl}\,\prime} =& -8 g^2 \gamma _c \gamma _h \gamma _m \left(\bar{n}_h \left(\bar{n}_c+\bar{n}_m+1\right)-\bar{n}_c \bar{n}_m\right) (3 \gamma _c \left(2 \bar{n}_c+1\right) \left(\gamma _h \left(2 \bar{n}_h+1\right)+\gamma _m \left(2 \bar{n}_m+1\right)\right)+\gamma _c^2 \left(2 \bar{n}_c+1\right){}^2\nonumber\\& ~+3 \gamma _h \gamma _m \left(2 \bar{n}_h+1\right) \left(2 \bar{n}_m+1\right)+\gamma _h^2 \left(2 \bar{n}_h+1\right){}^2+\gamma _m^2 \left(2 \bar{n}_m+1\right)^2),
\end{align}}
while the remanent coefficients are given in the repository \cite{SM_repository}.

\subsection{Noise-induced-coherent machine}

Finally, using the Hamiltonian and Lindblad operators from Eq. \eqref{eq: Hamiltonian NIC} and \eqref{eq:lindblad operators NIC}, we obtain the generic form of $W_G$ for the noise-induced-coherence machine. In contrast with the previous cases, now  we obtain non-zero real coherence in the steady state between states $\ket{\alpha}$ and $\ket{\beta}$(see App. \ref{appendix: classical equivalent NIC}). The relevant elements of the $W_G$ matrix then connect the populations of the four levels and the real part of the coherence between states $\ket{\alpha}$ and $\ket{\beta}$: 
\begin{equation*}
W_{G} = \left( \begin{array}{ccccc}
-(\gamma_{0\alpha}+\gamma_{01}) & \gamma_{10} ~\mathrm{e}^{-i\chi} & \gamma_{\alpha 0} & 0 & 0 \\
\gamma_{01}~\mathrm{e}^{i\chi} & -(\gamma_{1\alpha}+\gamma_{1\beta}+\gamma_{10}) & \gamma_{\alpha 1} & \gamma_{\beta 1} & 2\sqrt{\gamma_{\alpha 1}\gamma_{\beta 1}}\\
\gamma_{0\alpha} & \gamma_{1\alpha} & -(\gamma_{\alpha 1}+\gamma_{\alpha 0}) & 0 & - \sqrt{\gamma_{\alpha 1}\gamma_{\beta 1}}\\
0 & \gamma_{1\beta} & 0 & -\gamma_{\beta 1} & - \sqrt{\gamma_{\alpha 1}\gamma_{\beta 1}}\\
0 & 2\sqrt{\gamma_{1\alpha}\gamma_{1\beta}} & - \sqrt{\gamma_{\alpha 1}\gamma_{\beta 1}} & - \sqrt{\gamma_{\alpha 1}\gamma_{\beta 1}} & \gamma_{\alpha 0} + \gamma_{\alpha 1} + \gamma_{\beta 1}\end{array} \right),
\end{equation*}
with $\vec{p}_{G}(t) = (\rho_{00},\rho_{11},\rho_{\alpha\alpha},\rho_{\beta\beta},\mathrm{Re}[\rho_{\alpha\beta}])$. Some relevant coefficients of the inverse counting method are (all coefficients can be found in \cite{SM_repository}):
\begin{equation}
    \begin{split}
        &a_{0}^{\prime} = \gamma _{\text{eg}} \bar{n}_c \bar{n}_h \left(\bar{n}_h-\bar{n}_c\right) \left(\gamma _c{}^{\alpha }\right){}^2 \gamma _h{}^{\alpha } \gamma _h{}^{\beta }, ~~~~ a_{0}^{\prime\prime} = \gamma _{\text{eg}} \bar{n}_c \bar{n}_h \left(\bar{n}_c \left(2 \bar{n}_h+1\right)+\bar{n}_h\right) \left(\gamma _c{}^{\alpha }\right){}^2 \gamma _h{}^{\alpha } \gamma _h{}^{\beta },\\
        &a_{1} = -\gamma _c{}^{\alpha } \gamma _h{}^{\beta } \left(\gamma _{\text{eg}} \left(\bar{n}_c \left(4 \bar{n}_h+1\right)+\bar{n}_h\right) \left(\bar{n}_c \gamma _c{}^{\alpha }+\bar{n}_h \left(\gamma _h{}^{\alpha }+\gamma _h{}^{\beta }\right)\right)+\bar{n}_c \bar{n}_h \left(\bar{n}_c \left(4 \bar{n}_h+3\right)+3 \bar{n}_h+2\right) \gamma _c{}^{\alpha } \gamma _h{}^{\alpha }\right),\\
        &a_{1}^{\prime} = -\gamma _{\text{eg}} \left(\bar{n}_c-\bar{n}_h\right) \gamma _c{}^{\alpha } \gamma _h{}^{\alpha } \left(\bar{n}_c \gamma _c{}^{\alpha }+\bar{n}_h \gamma _h{}^{\alpha }\right).
    \end{split}
\end{equation}
The matrix for the classical equivalent of the NIC machine does not contain the coherence anymore, but the modified rates $\{\gamma_{1 \alpha}^\mathrm{cl}, \gamma_{\alpha 1}^\mathrm{cl}, \gamma_{1 \beta}^\mathrm{cl}, \gamma_{\beta 1}^\mathrm{cl}, \gamma_{\alpha \beta }^\mathrm{cl} \}$. It reads:
\begin{equation*}
W_{G}^{\mathrm{cl}} = \left( \begin{array}{ccccc}
-(\gamma_{0\alpha}+\gamma_{01}) & \gamma_{10}~\mathrm{e}^{-i\chi} & \gamma_{\alpha 0} & 0 \\
\gamma_{01}~\mathrm{e}^{i\chi} & -(\gamma_{1\alpha}^{\mathrm{cl}}+\gamma_{1\beta}^{\mathrm{cl}}+\gamma_{10}) & \gamma_{\alpha 1}^{\mathrm{cl}} & \gamma_{\beta 1}^{\mathrm{cl}} \\
\gamma_{0\alpha} & \gamma_{1\alpha}^{\mathrm{cl}} & -(\gamma_{\alpha 1}^{\mathrm{cl}}+\gamma_{\alpha 0}+\gamma_{\alpha\beta}^{\mathrm{cl}}) & \gamma_{\alpha\beta}^{\mathrm{cl}} \\
0 & \gamma_{1\beta}^{\mathrm{cl}} & \gamma_{\alpha\beta}^{\mathrm{cl}} & -(\gamma_{\beta 1}^{\mathrm{cl}}+\gamma_{\alpha\beta}^{\mathrm{cl}}) \end{array} \right),
\end{equation*}
 with associated vector $\vec{p}_{G}^{\,\mathrm{cl}}(t) = (\rho_{00},\rho_{11},\rho_{\alpha\alpha},\rho_{\beta\beta})$. Some relevant coefficients in this case are:
\begin{equation}
    \begin{split}
        &a_{0}^{\mathrm{cl}\,\prime} = \frac{\gamma _{\text{eg}} \bar{n}_c \bar{n}_h \left(\bar{n}_c-\bar{n}_h\right) \left(\gamma _c{}^{\alpha }\right){}^2 \gamma _h{}^{\alpha } \gamma _h{}^{\beta }}{\bar{n}_c \gamma _c{}^{\alpha }+\bar{n}_h \left(\gamma _h{}^{\alpha }+\gamma _h{}^{\beta }\right)}, ~~~~ a_{0}^{\mathrm{cl}\,\prime\prime} = \frac{\gamma _{\text{eg}} \bar{n}_c \bar{n}_h \left(\bar{n}_c-\bar{n}_h\right) \left(\gamma _c{}^{\alpha }\right){}^2 \gamma _h{}^{\alpha } \gamma _h{}^{\beta }}{\bar{n}_c \gamma _c{}^{\alpha }+\bar{n}_h \left(\gamma _h{}^{\alpha }+\gamma _h{}^{\beta }\right)},\\
        &a_{1}^{\mathrm{cl}} = \gamma _c{}^{\alpha } \gamma _h{}^{\beta } \left(\gamma _{\text{eg}} \left(\bar{n}_c \left(4 \bar{n}_h+1\right)+\bar{n}_h\right)+\frac{\bar{n}_c \bar{n}_h \left(\bar{n}_c \left(4 \bar{n}_h+3\right)+3 \bar{n}_h+2\right) \gamma _c{}^{\alpha } \gamma _h{}^{\alpha }}{\bar{n}_c \gamma _c{}^{\alpha }+\bar{n}_h \left(\gamma _h{}^{\alpha }+\gamma _h{}^{\beta }\right)}\right),\\
        &a_{1}^{\mathrm{cl}\,\prime} = \frac{\gamma _{\text{eg}} \left(\bar{n}_c-\bar{n}_h\right) \gamma _c{}^{\alpha } \gamma _h{}^{\alpha } \left(\bar{n}_c \gamma _c{}^{\alpha }+\bar{n}_h \left(\gamma _h{}^{\alpha }-\gamma _h{}^{\beta }\right)\right)}{\bar{n}_c \gamma _c{}^{\alpha }+\bar{n}_h \left(\gamma _h{}^{\alpha }+\gamma _h{}^{\beta }\right)},
    \end{split}
\end{equation}
the rest of them being given in the repository \cite{SM_repository}.

\section{Proof of Result I}
\label{appendix: proofs}

In this Appendix we provide the proof of Result 1 enunciated in the main text. The proof follows from the analytical evaluation of the sing of the ratio $\mathcal{R}$ in Eq. \eqref{eq:R}, which can be reduced to the evaluation of the difference between the variances in quantum and classical-equivalent machines, $\mathrm{Var}[\dot{N}^{\mathrm{cl}}] -  \mathrm{Var}[\dot{N}]$. Using the explicit expressions for the variances from the Inverse Counting Statistics method (see App. \ref{appendix: full counting}) provided in Eq.~\eqref{eq:var_with_coef}, that difference can be written as: 
\begin{equation}\label{eq:diference}
\begin{split}
    &\mathrm{Var}[\dot{N}^{\mathrm{cl}}] -  \mathrm{Var}[\dot{N}]= \underbrace{\left (\dfrac{a_{0}^{\prime\prime}}{a_1} - \dfrac{a_0^{cl\prime\prime}}{a_1^{cl}} \right)}_{(1)} \\ &+ 2 \langle\dot{N}\rangle\underbrace{\left( \dfrac{a_1^{\prime}}{a_1} - \dfrac{a_1^{cl\prime}}{a_1^{\mathrm{cl}}} \right)}_{(2)} + 2\langle\dot{N}\rangle^{2} \underbrace{\left(\dfrac{a_2}{a_1}-\dfrac{a_2^{cl}}{a_1^{cl}} \right)}_{(3)},
\end{split}
\end{equation}
where we took advantage from the fact that, by construction of the classical equivalent, $\langle \dot{N} \rangle = \langle \dot{N}^{\mathrm{cl}} \rangle$. In the following, we evaluate the three under-brace terms in Eq.~\eqref{eq:diference} for the inverse counting coefficients obtained for Hamiltonian-induced coherence [Eqs. \eqref{eq:terms_hamiltonian_induced} and \eqref{eq:terms_hamiltonian_induced_classical}]. We first consider the case of resonant driving $(\Delta_{\rm d} = 0)$ and then extend the analysis to non-zero detuning. For the first term $(1)$ we obtain always a vanishing contribution: 
\begin{equation}\label{eq:terms1}
    \left (\dfrac{a_{0}^{\prime\prime}}{a_1} - \dfrac{a_0^{cl\prime\prime}}{a_1^{cl}} \right) = 0,
\end{equation}
and therefore it does not contribute to Eq.~\eqref{eq:diference}. For the third term $(3)$ in Eq.~\eqref{eq:diference} we have instead:
\begin{equation}\label{eq:terms3}
\left(\dfrac{a_2}{a_1}-\dfrac{a_2^{cl}}{a_1^{cl}} \right) =  \dfrac{1}{\mathcal{Z}}\sum_{i\neq i^{\prime}}\sum_{j\neq j ^{\prime}}\sum_{k\neq k^{\prime}}\Gamma_{i i^{\prime}}\Gamma_{j j^{\prime}}\Gamma_{k k^{\prime}} \geq 0
\end{equation}
which is always positive since the rates are positive and $\mathcal{Z}>0$ is a positive (normalization) factor. Moreover, since this term is multiplied by $\langle \dot{N} \rangle^2$ in Eq.~\eqref{eq:diference}, it always leads to a positive contribution to $\mathrm{Var}[\dot{N}^{\mathrm{cl}}] -  \mathrm{Var}[\dot{N}]$.

On the other hand for the term $(2)$ in Eq.~\eqref{eq:diference} we have:
\begin{equation} \label{eq:terms2}
\begin{split}
 \left( \dfrac{a_1^{\prime}}{a_1} - \dfrac{a_1^{cl\prime}}{a_1^{\mathrm{cl}}} \right)& = \dfrac{1}{\mathcal{Z}}[\Gamma _{\text{vs}} \left(\Gamma _{\text{ms}} \Gamma _{\text{su}} \gamma _{\text{um}}-\gamma _{\text{mu}} \Gamma _{\text{us}} \Gamma _{\text{sm}} \right) \\ 
&+ \gamma_{\rm{vm}}(\gamma_{\rm{um}}\Gamma_{\rm{ms}}\Gamma_{\rm{su}}-\Gamma_{\rm{us}} \Gamma_{\rm{sm}} \gamma_{\rm{mu}}) \\ 
&+ (\gamma_{\rm{mu}}\Gamma_{\rm{us}}\Gamma_{\rm{sv}}\gamma_{\rm{vm}} - \gamma_{\rm{mv}}\Gamma_{\rm{vs}}\Gamma_{\rm{su}}\gamma_{\rm{um}})]
\end{split}
\end{equation}
where differences between the rates associated to certain cycles in opposite directions appear. However, in particular, none of these cycles involves the coherent transition between levels $\ket{u}$ and $\ket{v}$. Therefore, for unicyclic machines, we (trivially) have $\Gamma_{\rm{us}} = \Gamma_{\rm{su}} = 0$ and $\Gamma_{\rm{vm}} = \Gamma_{\rm{mv}} = 0$, which makes all the different terms in Eq.~\eqref{eq:terms2} equal to zero. Moreover, this term is also zero for symmetric multicycle quantum thermal machines, where all cycles have the same number of contributions from the different baths. This is the case, e.g. of multipartite systems under local dissipation. In that case we have that to close cycles that do not include the coherent transition between the levels $\ket{\rm u}$ and $\ket{\rm v}$, we will have the same number of quanta in and out from each given bath during the cycle (see e.g. Fig. 2b for an illustration) and therefore there will be no net probability flux in one or the other direction. This property implies that each of the three parenthesis in Eq.~\eqref{eq:terms2}, which are proportional to the conditional probabilities to close three of such cycles, cancels out.

As a consequence of the above analysis, we have that the fluctuation ratio $\mathcal{R}$ is given by the following expression: 
\begin{equation}
    \mathcal{R} = \dfrac{\langle \dot{N} \rangle^{2} }{\mathcal{Z}} \sum_{i\neq i^{\prime}}\sum_{j\neq j ^{\prime}}\sum_{k\neq k^{\prime}}\Gamma_{i i^{\prime}}\Gamma_{j j^{\prime}}\Gamma_{k k^{\prime}} \geq 0 
\end{equation}
which is always strictly greater that zero in out-of-equilibrium scenarios. The reason is that $\mathcal{R}$ becomes zero only in the case of $\langle\dot{N} \rangle = 0$, which corresponds to the equilibrium point.

To analyze the effect of detuning in the driving, we now repeat the same steps as in the earlier demonstration but using the more general expressions derived within this scenario. We find that, as before, \eqref{eq:terms1} does not contribute to $\mathcal{R}$. However, in this case, \eqref{eq:terms3} and \eqref{eq:terms2} do not consistently exhibit a well-defined positive sign in all cases. By considering the types of uni-cyclic and multi-cyclic machines to which our result applies, it can be shown that the fluctuation ratio takes the form
\begin{equation}\label{eq:detu}
    \mathcal{R} = \left(\left[\dfrac{1}{2}\sum_{i}(\gamma_{\rm u i}+\gamma_{\rm v i})\right]^2-\Delta_{\rm d}^2\right)\Lambda,
\end{equation}
where $\Lambda \geq 0$ represents a sum of rates (for its explicit expression see Ref.~\cite{SM_repository}). As a consequence, we have $\mathcal{R} \geq 0$ whenever the expression within parentheses in Eq.~\eqref{eq:detu}. Result 1 remains thus valid in the presence of detuning when $|\Delta_{\rm d}| < \dfrac{1}{2}\sum_{i}(\gamma_{\rm u i}+\gamma_{\rm v i})$ is verified. 

\section{Details on the NIC machine analysis}
\label{appendix: NICdetails}
In this Appendix we discuss in more detail the change of variables introduced in Sec.~\ref{sec:illustrativeC} to decouple one of the two collective transitions of the NIC coherent machine, together with extra comments regarding the symmetric limit in which all the spontaneous emission rates are equal. 

\begin{figure}
  \centering\includegraphics[width=0.8\textwidth]{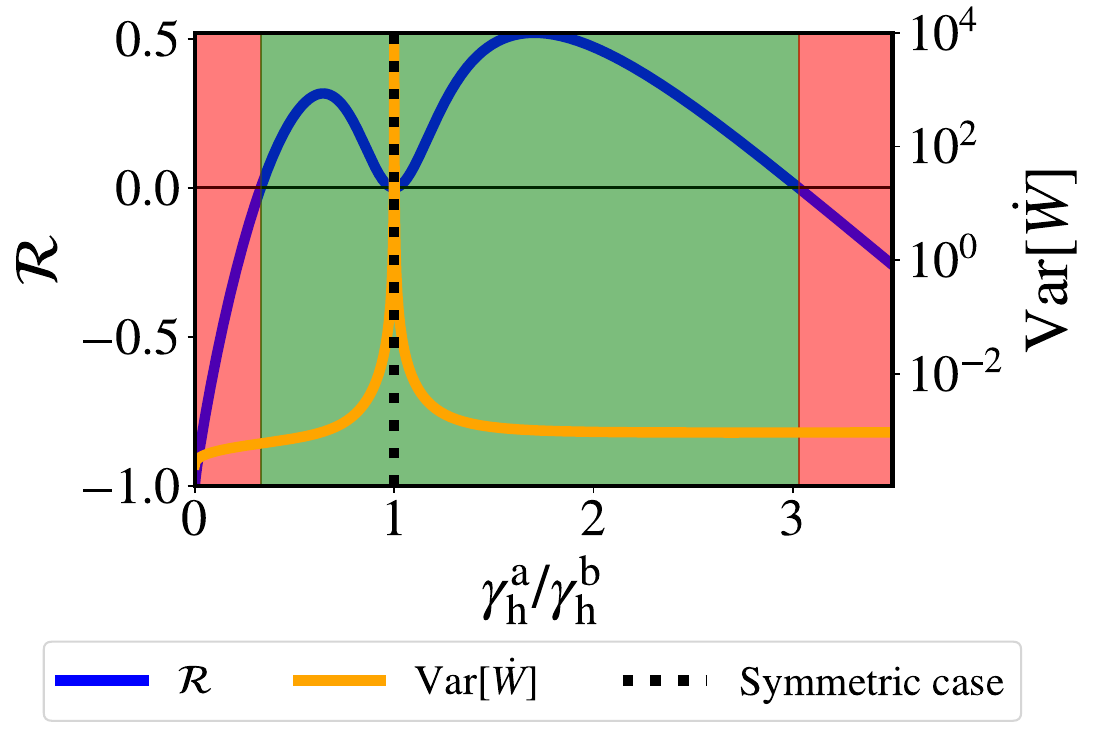}
  \caption{Fluctuations ratio $\mathcal{R}$ (blue solid line) for the NIC machine and variance in the quantum system (orange doted line) as functions of the ratio of the original spontaneous emission rates (prior to the change of variables). The system parameters are:  $\beta_{\mathrm{w}} \rightarrow 0$, $\beta_{\mathrm{c}} = 1$, $\beta_{\mathrm{h}}/\beta_{\mathrm{c}} = 0.9$, $\epsilon_{2} = 5$, $\epsilon_{1} = 3$, and $\gamma_{\mathrm{c}}^{(\mathrm{a}/\mathrm{b})} = 10^{-3}$ (energies are given in units of $k_B T_{\rm c} = 1$).}
   \label{fig:symmetric_case}
\end{figure}
Once we have introduced the change of variable \eqref{eq: variable_change} we can rewrite the set of new Lindblad operators as  
\begin{equation}\label{}
    \begin{split}
        L_{\downarrow}^{(\text{c})} =&  \sqrt{\gamma_{\alpha 0}}\ket{0}\bra{\alpha},\;
        L_{\uparrow}^{(\text{c})} = \sqrt{\gamma_{0\alpha}}\ket{\alpha}\bra{0},\\
        L_{\downarrow}^{(\text{h})} &= \sqrt{\gamma_{\alpha 1}}\ket{1}\bra{\alpha} + \sqrt{\gamma_{\beta 1}}\ket{1}\bra{\beta}  ,\\
        L_{\uparrow}^{(\text{h})} &= \sqrt{\gamma_{1\alpha}}\ket{\alpha}\bra{1} + \sqrt{\gamma_{1\beta}}\ket{\beta}\bra{1},  
    \end{split}
\end{equation}
where the different transitions are mediated by the same thermal baths as before, i.e $\gamma_{\alpha 0} = (\bar{n}_{\mathrm{c}}+1)\gamma_{\mathrm{c}}^{\alpha}$, $\gamma_{0\alpha} = \bar{n}_{\mathrm{c}}\gamma_{\mathrm{c}}^{\alpha}$, $\gamma_{\alpha 1} = (\bar{n}_{\mathrm{h}}+1)\gamma_{\mathrm{h}}^{\alpha}$, $\gamma_{1\alpha} = \bar{n}_{\mathrm{h}}\gamma_{\mathrm{h}}^{\alpha}$, $\gamma_{\beta 1} = (\bar{n}_{\mathrm{h}}+1)\gamma_{\mathrm{h}}^{\beta}$, $\gamma_{1\beta} = \bar{n}_{\mathrm{h}}\gamma_{\mathrm{h}}^{\beta}$; and we can identify the new spontaneous emission rates:
\begin{equation}
\begin{split}
\gamma_{\mathrm{c}}^{\alpha} &= \gamma_{\mathrm{c}}^{\mathrm{a}}+\gamma_{\mathrm{c}}^{\mathrm{b}}, \\
\gamma_{\mathrm{h}}^{\alpha} &= (\sqrt{\gamma_{\mathrm{c}}^{\mathrm{a}}\gamma_{\mathrm{h}}^{\mathrm{a}}}+\sqrt{\gamma_{\mathrm{c}}^{\mathrm{b}}\gamma_{\mathrm{h}}^{\mathrm{b}}})^{2}/(\gamma_{\mathrm{c}}^{\mathrm{a}}+\gamma_{\mathrm{c}}^{\mathrm{b}}), \\
\gamma_{\mathrm{h}}^{\beta} &= (\sqrt{\gamma_{\mathrm{c}}^{\mathrm{a}}\gamma_{\mathrm{h}}^{\mathrm{b}}}-\sqrt{\gamma_{\mathrm{c}}^{\mathrm{b}}\gamma_{\mathrm{h}}^{\mathrm{a}}})^{2}/(\gamma_{\mathrm{c}}^{\mathrm{a}}+\gamma_{\mathrm{c}}^{\mathrm{b}}).
\end{split}
\end{equation}

For symmetric rates, we recover the situation where $\ket{\beta}$ is completely decoupled from the states $\ket{0}$ and $\ket{1}$ ($\gamma_{\mathrm{h}}^{\beta}=0$), leading to a classical three-level system with local jumps induced by thermal baths at same temperatures than the original NIC machine, plus an extra uncoupled level (dark state). Considering Eqs.~\eqref{eq:rates_noise-induced}, \eqref{eq:newrates} and \eqref{eq: validity of the equivalent}, it becomes evident that in this limit, the quantum system and its classical analog become indistinguishable once the probability of the independent (dark) level is set to zero. On the other hand, if we approach this symmetric case asymptotically, the first inequality in Eq. ~\eqref{eq:conditions_amplification} indicates that, in principle, the quantum system is more precise than the classical equivalent. However, it is important to note that in this limit, fluctuations in the system currents will diverge due to a first-order phase transition, as shown in Ref.~\cite{Holubec19}. That leads to $\mathcal{R}=0$ as can be appreciated in Fig.~\ref{fig:symmetric_case} for $\gamma_{\rm h}^{\rm a} = \gamma_{\rm h}^{\rm b}$, where ${\rm Var}[\dot{W}]$ diverges (vertical dotted line). The green and red areas in Fig.~\ref{fig:symmetric_case} represent, respectively, the regions of parameters where quantum advantages or disadvantages are obtained. This finding aligns with the observation that, in the symmetric case, the change of variables completely decouples the coherences from the populations. As mentioned before, in that case one can always redefine the relevant thermal machine under consideration by excluding the independent (dark) level. In that case, the remaining (connected) thermal machine will be exactly equal as their classical counterpart.

\end{document}